\documentclass[prd, 11pt,nofootinbib, superscriptaddress, preprintnumbers,floatfix]{revtex4}
\usepackage{amssymb,amsmath}
\usepackage{graphicx}
\usepackage{multirow}
\usepackage{comment}
\usepackage{color}

\preprint{JLAB-THY-09-1060}
\preprint{NT@UW 09-17}
\begin{document}
\newcommand{\ignore}[1]{}
\def\mc#1{{\mathcal #1}}
\def\L{\Lambda}
\def\D{\Delta}
\def\d{\delta}
\def\a{\alpha}
\def\b{\beta}
\def\S{\Sigma}
\def\s{\sigma}
\def\e{\epsilon}
\def\O{\Omega}
\def\k{\kappa}
\def\g{\gamma}

\title{Singly and Doubly Charmed $J=1/2$ Baryon Spectrum from Lattice QCD}

\author{Liuming Liu}
\email{lxliux@wm.edu}
\affiliation{Department of Physics, College of William, Williamsburg, VA 23187-8795}
\affiliation{Thomas Jefferson National Accelerator Facility, Newport News, VA 23606}

\author{Huey-Wen Lin}
%\email{hwlin@phys.washington.edu}
\affiliation{Thomas Jefferson National Accelerator Facility, Newport News, VA 23606}
\affiliation{Department of Physics, University of Washington, Seattle, WA 98195-1560}

\author{Kostas Orginos}
%\email{kostas@wm.edu}
\affiliation{Department of Physics, College of William, Williamsburg, VA 23187-8795}
\affiliation{Thomas Jefferson National Accelerator Facility, Newport News, VA 23606}

\author{Andr\'{e} Walker-Loud}
%\email{walkloud@wm.edu}
\affiliation{Department of Physics, College of William, Williamsburg, VA 23187-8795}

\begin{abstract}
We compute the masses of the singly and doubly charmed baryons in full QCD using the relativistic Fermilab action for the charm quark. For the light quarks we use domain-wall fermions in the valence sector and improved Kogut-Susskind sea quarks. We use the low-lying charmonium spectrum to tune our heavy-quark action and as a guide to understanding the discretization errors associated with the heavy quark. 
Our results are in good agreement with experiment within our systematics.
For the $\Xi_{cc}$,  we find the isospin-averaged mass to be $M_{\Xi_{cc}} = 3665 \pm17 \pm14\, {}^{+0}_{-78}$~MeV; the three given uncertainties are statistical, systematic and an estimate of lattice discretization errors, respectively.
In addition, we predict the mass splitting of the (isospin-averaged) spin-1/2 $\O_{cc}$ with the $\Xi_{cc}$ to be $M_{\O_{cc}} - M_{\Xi_{cc}} = 98 \pm9 \pm22\pm13$~{MeV} (in this mass splitting, the leading discretization errors are also suppressed by $SU(3)$ symmetry).
Combining this splitting with our determination of $M_{\Xi_{cc}}$ leads to our prediction of the spin-1/2 $\O_{cc}$ mass, $M_{\O_{cc}} = 3763\pm19\pm26\,{}^{+13}_{-79}$~{MeV}.
\end{abstract}

%%%%%%%%%%%%%%%%%%%%%%%%%%%%%%%%%%%%%%%%%%%%%%%%%%%%%%%%%%%%%%%%%%%%%%%%%%%%%%%%%%%%%%
\maketitle

%\newpage
%\tableofcontents

%%%%%%%%%%%%
%
%	INTRODUCTION
%
%%%%%%%%%%%%
\section{Introduction}
Experimental and theoretical studies of charmed and bottom hadrons have been the focus of vigorous research over the last several years~\cite{Barberio:2008fa,Voloshin:2007dx,:2007rw,:2007ub}. In particular, singly and doubly heavy baryon spectroscopy has received significant attention, mainly due to the recent experimental discoveries of both new charmed (SELEX)~\cite{Mattson:2002vu,Ocherashvili:2004hi} and bottom baryons by D0~\cite{:2007ub} and CDF~\cite{:2007un}. In addition to these discoveries, there are still many states of heavy and doubly heavy baryons remaining to be discovered.
The new Beijing Spectrometer (BES-III), a detector at the recently upgraded Beijing Electron Positron Collider (BEPCII), has great potential for accumulating large numbers of events to help us understand more about charmed hadrons.
The antiProton ANnihilation at DArmstadt (PANDA) experiment, a GSI future project, and the LHCb are also expected to provide new results to help experimentally map out the heavy-baryon sector. For these reasons, lattice quantum chromodynamic (QCD) calculations of the spectrum of heavy baryons are now very timely and will play a significant role in providing theoretical first-principles input to the experimental program.

Lattice QCD is now a mature field capable of providing accurate results that can be directly compared to experiment, with  calculations in the light-quark sector being well established. Although  the study of heavy quarks requires careful treatment of discretization errors, significant advances have been made in this sector as well. Lattice heavy quarks have $\mc{O}((m_Q\,a)^n)$ errors, where $m_Q$ is the mass of the heavy quark and $a$ is the lattice spacing. Lattice spacings for typical, currently accessible dynamical ensembles are still too coarse ($a^{-1} \approx2\mbox{ GeV}$) to make such systematic errors small.
To assert better control over the discretization errors for heavy quarks on the lattice, several heavy-quark approaches have proven useful.
For example, non-relativistic QCD (NRQCD)~\cite{Lepage:1992tx}, which is an expansion of the lattice quark action in powers of $\frac{1}{am_Q}$, is commonly applied to bottom quarks. However, the  charm-quark mass is not heavy enough to justify the use of NRQCD.
Relativistic heavy-quark actions~\cite{ElKhadra:1996mp,Aoki:2001ra,Christ:2006us,Lin:2006ur} systematically remove $\mc{O}((m_Qa)^n)$ terms and are better suited to charm-quark calculations. Recent updates on the state of heavy-quark physics on the lattice can be found in several reviews~\cite{Kronfeld:2003sd,Wingate:2004xa,Okamoto:2005zg,Onogi:2006km,DellaMorte:2007ny,Gamiz:2008iv} and references therein.

Up to now, there have been a few lattice charmed-baryon calculations using the quenched approximation. In some cases an $\mc{O}(a)$-improved light-quark action is used on isotropic or anisotropic lattices with a single lattice spacing:
Bowler et~al.~\cite{Bowler:1996ws} used a tree-level clover action for both light and heavy quarks to calculate the singly charmed baryons spectrum of spin 1/2 and 3/2.
Later, Flynn et~al.~\cite{Flynn:2003vz} updated this project with nonperturbative clover action and extended the calculation to doubly charmed baryons.
Chiu et~al.~\cite{Chiu:2005zc} used a chiral fermion action for the charm quarks and calculated both the positive and negative parity spectrum for singly and doubly charmed baryons.
Such calculations using light-quark actions to simulate heavy quarks introduce large systematic errors proportional to $(am_Q)^2$, which must be carefully addressed.
One calculation has used a higher-order improved fermion action:
Lewis et~al.~\cite{Lewis:2001iz} performed a calculation on both doubly and singly charmed baryons using D234-type fermion action (which would leave a leading error of $\mc{O}(a^3)$) for both light and heavy quarks but on a coarse anisotropic ensemble (with anisotropy $\xi=2$).
Finally, heavy-quark effective theory was applied to charm calculation:
Mathur et~al.~\cite{Mathur:2002ce} continued to use anisotropic lattices, adding two more lattice spacings, but changed the heavy-quark action to NRQCD, which reduces the lattice-spacing discretization effects.
For all of these calculations, the quenched approximation remains a significant source of systematic error that is difficult to estimate.

Given the progress on the experimental side, it is time to revisit these charmed baryon calculations using dynamical gauge ensembles and improve the calculations with the current available computational resources.
Although more dynamical ensembles are available these days, not many charmed baryon calculations have been published so far, only a few proceedings~\cite{Na:2007pv,Na:2008hz,Liu:2008rza}.

In this work, we extend our previous calculation~\cite{Liu:2008rza} to higher statistics and compute the ground-state spectrum of the spin-1/2 singly and doubly charmed baryons. We use the Fermilab action~\cite{ElKhadra:1996mp} for the charm quarks and domain-wall fermions for the light valence quarks on gauge configurations with 2+1-flavor Kogut-Susskind fermions and a range of quark masses resulting in pion masses as light as 290~{MeV}.
We nonperturbatively tune the fermion anisotropy and two input bare masses for charm quarks, setting the remaining parameters to tree-level tadpole improved coefficients.  Our results are extrapolated to the physical light-quark masses using both heavy-hadron chiral perturbation theory (HH$\chi$PT) as well as HH$\chi$PT-inspired polynomial extrapolations.

%%%%%%%%%%%%
%
%		LATTICE ACTION
%
%%%%%%%%%%%%
\section{Lattice Formulation\label{sec:LatticeAction}}

%%%%%%%%%%%%
%
%		LIGHT-QUARK ACTION
%
%%%%%%%%%%%%
\subsection{Light-Quark Action\label{sec:lQaction}}
In this work we employ the ``coarse'' ($a\simeq 0.125$~fm) gauge configurations generated by the MILC Collaboration~\cite{Bernard:2001av} using the one-loop tadpole-improved gauge action~\cite{Alford:1995hw}, where both $\mc{O}(a^2)$ and $\mc{O}(g^2a^2)$ errors are removed. For the fermions in the vacuum, the asqtad-improved Kogut-Susskind action~\cite{Orginos:1999cr,Orginos:1998ue,Toussaint:1998sa,Lagae:1998pe,Lepage:1998vj,Orginos:1999kg} is used. This is the Naik action~\cite{Naik:1986bn} ($\mc{O}(a^2)$ improved Kogut-Susskind action) with smeared links for the one-link terms so that couplings to gluons with any of their momentum components equal to $\pi/a$ are set to zero.

For the valence light quarks (up, down and strange) we use the five-dimensional Shamir~\cite{Shamir:1993zy,Furman:1994ky} domain-wall fermion propagators~\cite{Kaplan:1992bt} calculated by the NPLQCD Collaboration~\cite{Beane:2008dv}. The domain-wall fermion action introduces a fifth dimension of extent $L_5$ and a mass parameter $M_5$; in our case the values $L_5=16$ and $M_5=1.7$ were chosen. The physical quark fields, $q(\vec x, t)$, reside on the 4-dimensional boundaries of the fifth coordinate. The left and right chiral components are separated on the corresponding boundaries, resulting in an action with chiral symmetry at finite lattice spacing as $L_5 \rightarrow \infty$. We use hypercubic-smeared gauge links~\cite{Hasenfratz:2001hp,DeGrand:2002vu,DeGrand:2003in,Durr:2004as} to minimize the residual chiral symmetry breaking, and the bare quark-mass parameter $(a m)^{\rm dwf}_q$ is introduced as a direct coupling of the boundary chiral components.

The calculation we have performed, because the valence and sea quark actions are different, is inherently partially quenched and therefore violates unitarity.  Unlike conventional partially quenched calculations, to restore unitarity, one must take the continuum limit in addition to tuning the valence and sea quark masses to be degenerate.  This process is aided with the use of mixed-action chiral perturbation theory~\cite{Bar:2005tu,Tiburzi:2005is,Chen:2006wf,Orginos:2007tw,Chen:2007ug,Chen:2009su}. Given the situation, there is an ambiguity in the choice of the valence light-quark masses.  One appealing choice is to tune the masses such that the valence pion mass is degenerate with one of the staggered pion masses.  In the continuum limit, the $N_f=2$ staggered action has an $SU(8)_L\otimes SU(8)_R\otimes U(1)_V$ chiral symmetry due to the four-fold taste degeneracy of each flavor, and each pion has 15 degenerate partners. At finite lattice spacing this symmetry is broken and the taste multiplets are no longer degenerate, but have splittings that are $\mc{O}(\alpha_s^2 a^2)$~\cite{Orginos:1999cr,Orginos:1998ue,Toussaint:1998sa,Orginos:1999kg,Lee:1999zxa}.  The propagators used in this work were tuned to give valence pions that match the Goldstone Kogut-Susskind pion. This is the only pion that becomes massless in the chiral limit at finite lattice spacing. As a result of this choice, the valence pions are as light as possible, while being tuned to one of the staggered pion masses, providing better convergence in the $\chi$PT needed to extrapolate the lattice results to the physical quark-mass point.  This set of parameters, listed in Table~\ref{table:configuration}, was first used by LHPC~\cite{Renner:2004ck,Edwards:2005kw} and recently to compute the spectroscopy hadrons composed of up, down and strange quarks~\cite{WalkerLoud:2008bp}.

\begin{table}
\begin{ruledtabular}
\begin{tabular}{c|ccccccc}
Ensemble      &$\beta$ &$am_l$    &$am_s$   & $am_l^{\rm dwf}$ & $am_s^{\rm dwf}$ & $N_{\rm cfgs}$ &$N_{\rm props}$ \\
\hline
\texttt{m007}& 6.76   & 0.007    & 0.050   & 0.0081   & 0.081    & 461   & 2766 \\
\texttt{m010}& 6.76   & 0.010    & 0.050   & 0.0138   & 0.081    & 636   & 3816 \\
\texttt{m020}& 6.79   & 0.020    & 0.050   & 0.0313   & 0.081    & 480   & 1920 \\
\texttt{m030}& 6.81   & 0.030    & 0.050   & 0.0478   & 0.081    & 563   & 1689 \\
\end{tabular}
\end{ruledtabular}
\caption{\label{table:configuration}The parameters of the configurations and domain-wall propagators used in this work. The subscript $l$ denotes light quark, and $s$ denotes the strange quark. The superscript ``dwf'' denotes domain-wall fermion.}
\end{table}

%%%%%%%%%%%%
%
%		HEAVY-QUARK ACTION
%
%%%%%%%%%%%%
\subsection{Heavy-Quark Action\label{sec:HQaction}}
For the charm quark we use the Fermilab action~\cite{ElKhadra:1996mp}, which controls discretization errors of $\mc{O}((a m_Q)^n)$.
Following the Symanzik improvement~\cite{Symanzik:1983dc}, an effective continuum action is constructed using operators that are invariant under discrete rotations, parity-reversal and charge-conjugation transformations, representing the long-distance limit of our lattice theory, including leading finite-$a$ errors.
Using only the Dirac operator and the gluon field tensor (and distinguishing between the time and space components of each), we enumerate seven operators with dimension up to five.
By applying the isospectral transformations~\cite{Chen:2000ej}, the redundant operators are identified and their coefficients are set to appropriate  convenient values.
The lattice action then takes the form
\begin{equation}\label{eq:SFermi}
	S = S_0+S_B+S_E\, ,
\end{equation}
with
\begin{align}
\label{eq:action_S_0}
&S_0 = \sum_x \bar{Q}(x)\left[m_0+\left(\gamma_0 \nabla_0-\frac{a}{2}\triangle_0\right)
	+ \nu \sum_i\left(\gamma_i\nabla_i-\frac{a}{2}\triangle_i\right)
	\right]Q(x)\, ,& \\
\label{eq:action_S_B}
&S_B = -\frac{a}{2}c_B\sum_x \bar{Q}(x)\left(\sum_{i<j}\sigma_{ij}F_{ij} \right) Q(x)\, ,&\\
\label{eq:action_S_E}
&S_E = -\frac{a}{2}c_E\sum_x \bar{Q}(x) \left(\sum_i\sigma_{0i}F_{0i} \right) Q(x)\, ,&
\end{align}
where $a$ is the lattice spacing, $\nabla_0$ and $\nabla_i$ are first-order lattice derivatives in the time and space directions, $\triangle_0$ and $\triangle_i$ are second-order lattice derivatives, and $F_{\mu\nu}$ is the gauge field strength tensor.
The spectrum of heavy-quark bound states can be determined accurately through $|\vec{p}|a$ and $(am_Q)^n$ for arbitrary exponent $n$ by using a lattice action containing $m_0$, $\nu$, $c_B$ and $c_E$, which are functions of $am_Q$.

The coefficients $c_B$ and $c_E$ are different due to the broken space-time interchange symmetry, which can be computed in perturbation theory by requiring elimination of the heavy-quark discretization errors at a given order in the strong coupling constant $\alpha_s$. We use the tree-level tadpole-improved results obtained by using field transformation (as in Ref.~\cite{Chen:2000ej}):
\begin{equation}\label{eq:cBcE}
	c_B=\frac{\nu}{u_0^3}, \quad \quad
	c_E=\frac{1}{2}(1+\nu)\frac{1}{u_0^3} ,
\end{equation}
where $u_0$ is the tadpole factor
\begin{equation}
	u_0 = \left \langle \frac{1}{3} \sum_p {\rm Tr} (U_p )\right\rangle^{1/4}\;,
\end{equation}
and $U_p$ is the product of gauge links around the fundamental lattice plaquette $p$.
The remaining two parameters $m_0$ and $\nu$ are determined nonperturbatively. The bare charm-quark mass $m_0$ is tuned so that the experimentally observed spin average of the $J/\Psi$ and $\eta_c$ masses
\begin{equation}
	M_{\rm avg} = \frac{1}{4} M_{\eta_c} + \frac{3}{4} M_{J/\Psi}
\label{eq:spin-av}
\end{equation}
is reproduced; see Sec.~\ref{sec:HQactionTest} for further details.  The value of $\nu$ must be tuned to restore the dispersion relation $E_h^2=m_h^2+c^2p^2$ such that $c^2 = 1$.  Since the values of $\nu$ and $m_0$ are coupled, one needs to iterate the tuning process in order to achieve a consistent pair of values.  
To do this, we calculate the single-particle energy of $\eta_c$, $J/\Psi$, $D_s$ and $D$ at the six lowest momenta (with unit of $a^{-1}$): $\frac{2\pi}{L}(0,0,0), \frac{2\pi}{L}(1,0,0),$ $ \frac{2\pi}{L}(1,1,0),\frac{2\pi}{L}(1,1,1), \frac{2\pi}{L}(2,0,0), \frac{2\pi}{L}(2,1,0)$. For each ensemble, the energy levels are calculated at two charm-quark masses (denoted $m_1=0.2034$ and $m_2=0.2100$) and extrapolated to the physical charm-quark mass (as described below). The values of $c^2$ are obtained by fitting the extrapolated energy levels to the dispersion relation.  We tune $\nu$ using the dispersion relation of $\eta_c$.  As one can see from Table~\ref{tab:C}, the dispersion relations for either the charmonium $J/\Psi$ or the charm-light mesons ($D$ and $D_s$) are generally consistent with $c^2 = 1$ to within 1-2\%.
%%%%%%%%%%%%%%%%%%%%%%%
%		TABLE: SPEED OF LIGHT
%%%%%%%%%%%%%%%%%%%%%%%
\begin{table}
\begin{ruledtabular}
\begin{tabular}{c|cccc}
& \multicolumn{4}{c}{$c^2$} \\
\cline{2-5}
Ensemble   & $\eta_c$           & $J/\Psi$                       &$D$                            &$D_s$         \\
\hline
    \texttt{m007}            & $0.991(4)$    &$0.985(5)$     &$1.021(15)$   &$1.018(9)$   \\
   % \hline
    \texttt{m010}           &$0.989(3) $       &$0.958(3)$    &$1.016(10)$     &$0.992(6)$ \\
   % \hline
    \texttt{m020}           &$0.997(4)$      &$0.993(5)$     &$1.019(20)$    &$1.004(14)$  \\
   % \hline
    \texttt{m030}            &$0.963(5)$      &$0.947(6)$         &$1.029(12)$        &$1.015(10)$ \\
\end{tabular}
\end{ruledtabular}
\caption{ \label{tab:C}   Speed of light for charmed mesons.}
\end{table}
%%%%%%%%%%%%%%%%%%%%%%%

%%%%%%%%%%%%
%
%		CHARMED SPECTRUM RESULTS
%
%%%%%%%%%%%%
\section{Charmed Hadron Spectrum: Numerical Results\label{sec:spectrum}}

The interpolating operators we use for the $J=1/2$ singly and doubly charmed baryons are
\begin{eqnarray}
	\Lambda_{c}  :  && \epsilon^{ijk}(q_u^{iT} C\gamma_5 q_d^j)Q_c^k, \nonumber\\
	\Xi_{c}  :  && \epsilon^{ijk}(q_u^{iT} C\gamma_5 q_s^j)Q_c^k, \nonumber\\
	\Sigma_{c}  :  && \epsilon^{ijk}(q_u^{iT} C\gamma_5 Q_c^j)q_u^k, \nonumber\\
	\Xi^\prime_{c}  :  &&\frac{1}{\sqrt{2}} \epsilon^{ijk}\left[(q_u^{iT} C\gamma_5 Q_c^j)q_s^k
		+ (q_s^{iT} C\gamma_5 Q_c^j)q_u^k\right] , \nonumber\\
	\Omega_{c}  :  && \epsilon^{ijk}(q_s^{iT} C\gamma_5 Q_c^j)q_s^k, \nonumber\\
	\Xi_{cc}  :  && \epsilon^{ijk}(Q_c^{iT} C\gamma_5 q_u^j)Q_c^k, \nonumber\\
	\Omega_{cc}  : &&  \epsilon^{ijk}(Q_c^{iT} C\gamma_5 q_s^j)Q_c^k,
\end{eqnarray}
where $q_{u,d}$ are the up and down quark fields, $q_s$ is strange quark field and $Q_c$ is charm quark field.

%\subsection{Charmed Hadron Spectrum\label{sec:spectrum}}
Using these interpolating fields, we construct the two-point functions
\begin{equation}
	C_h(t,t_0)=\sum_\mathbf{x}\langle \mathcal{O}_h(\mathbf{x},t)\mathcal{O}_h(\mathbf{x},t_0)^\dagger\rangle  , \nonumber
\end{equation}
where $\mathcal{O}_h$ is an interpolating operator of the hadron $h$.
The correlation functions are calculated with gauge-invariant Gaussian-smeared sources and point sinks.  The smearing parameters were optimized so that excited-state contamination to the correlators is  minimized.
The domain-wall valence propagators were computed with Dirichlet boundary conditions in the time direction, reducing the original lattices to half their temporal size.  Similar to baryons, the signal for the charmed correlation functions quickly drops, and thus we do not expect the temporal reduction to reduce the number of useful time points for our analysis.  The sources were located away from the Dirichlet boundary to minimize contamination from the boundary effects.  In order to enhance our statistical precision, several valence propagators are taken from each configuration with varying source location.  The resulting correlation functions are then source averaged on each configuration to produce one correlator per configuration for each interpolating operator.  The masses of the hadrons are obtained by fitting the correlation functions to a single exponential
\begin{equation}
	C_h(t)=Ae^{-E_0t}
\end{equation}
in a region where the effective mass is observed to exhibit a plateau.  The fitting range is varied to estimate the systematics from the choice of fitting window, as indicated in Tables~\ref{tab:baryons} and \ref{tab:mesons}.

\begin{figure}
\begin{tabular}{cc}
\includegraphics[width=0.45\textwidth]{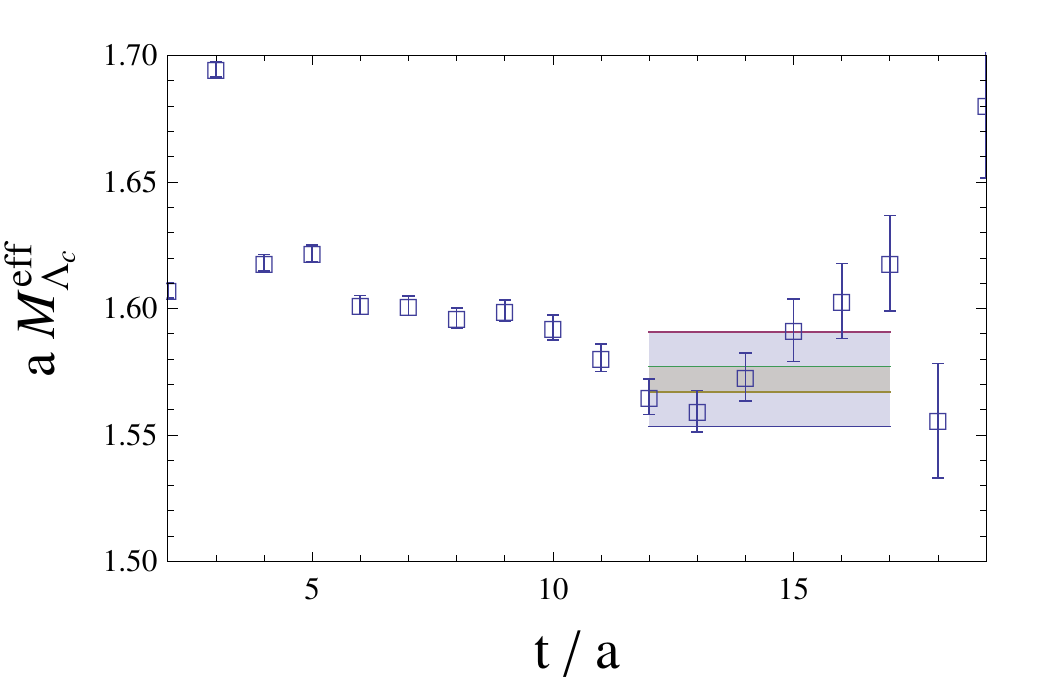}
&
\includegraphics[width=.45\textwidth]{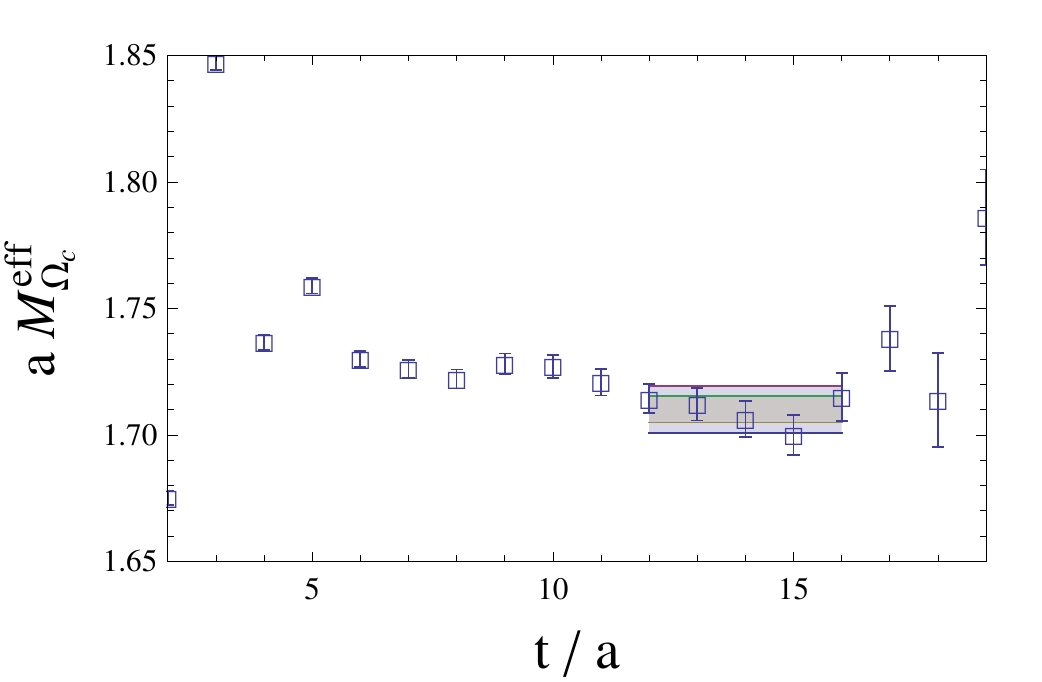}
\\
\includegraphics[width=0.45\textwidth]{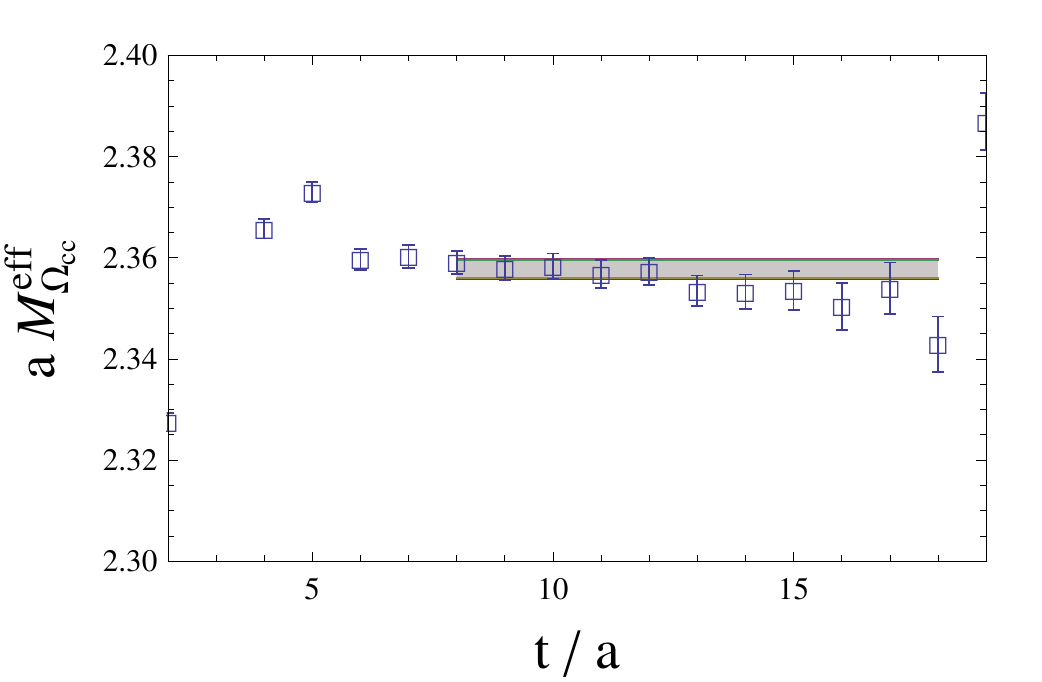}
&
\includegraphics[width=0.45\textwidth]{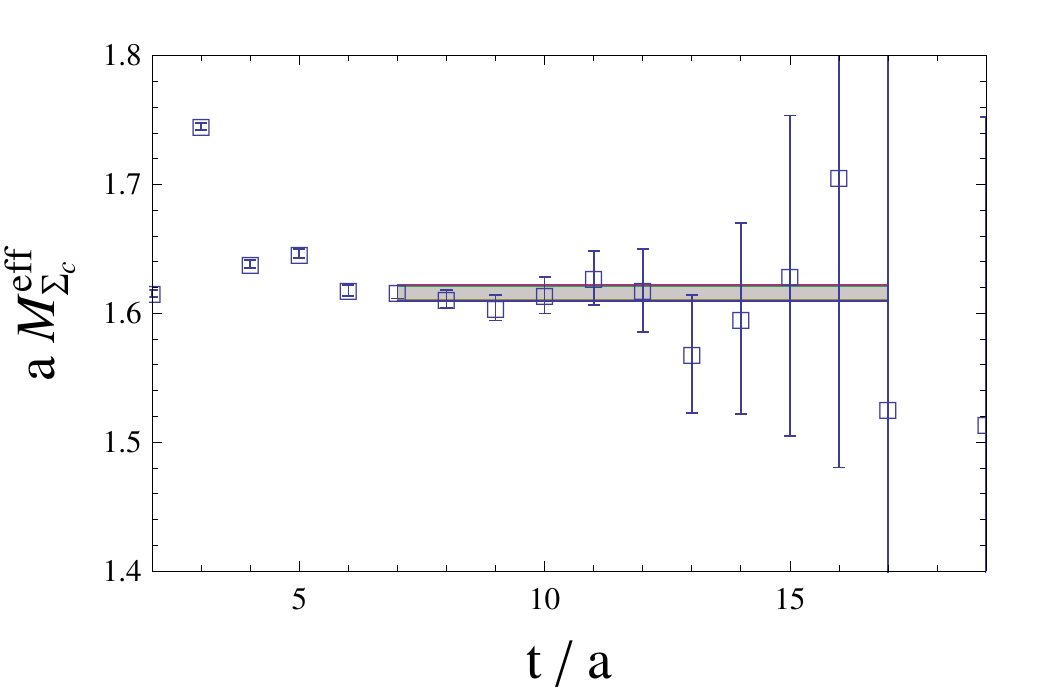}
\end{tabular}
\caption{ \label{fig:effctive_mass}Sample effective-mass plots and corresponding fits to the correlation functions.  The smaller error bands are statistical and the larger error bands are statistical and systematic (determined by varying fit range) added in quadrature.}
\end{figure}

In Table~\ref{tab:baryons}, we summarize the resulting baryon masses as well as the corresponding time ranges.  The first uncertainty is statistical and the second is a fitting systematic.  For most fits, the resulting $\chi^2$ per degree of freedom is  about one.  In Figure~\ref{fig:effctive_mass} we display representative effective mass plots and their fitted masses for both good and poor fits.  The results from charmonium are shown in Table~\ref{tab:mesons}.

%%%%%%%%%%%%%%%%
%%%  Results
%%%%%%%%%%%%%%%%
\begin{table}[tb]
\begin{ruledtabular}
\begin{tabular}{c|cllll}
Hadron& $m_0$& \texttt{m007}& \texttt{m010} &\texttt{m020}& \texttt{m030}\\
%\multicolumn{2}{c}{\texttt{m007}}& \multicolumn{2}{c}{\texttt{m010}}& \multicolumn{2}{c}{\texttt{m020}}& \multicolumn{2}{c}{\texttt{m030}}\\
%\cline{3}\cline{4-5}\cline{6-7}\cline{8-9}
%&$\k_1$& $\k_2$ &$\k_1$& $\k_2$ &$\k_1$& $\k_2$ &$\k_1$& $\k_2$\\
\hline
$\O_{cc}$& $m_1$& 2.3578(18)(8)[8--18]& 2.3620(14)(9)[10--18]& 2.3456(33)(17)[12--18]& 2.3333(23)(6)[11--18]\\
& $m_2$& 2.3663(18)(8)[8--18]& 2.3705(14)(9)[10--18]& 2.3542(33)(16)[12--18]& 2.3419(23)(7)[11--18]\\
\hline
%%% Xi
$\Xi_{cc}$& $m_1$& 2.3018(27)(0)[7--13]& 2.3120(23)(23)[9--17]& 2.3087(33)(3)[8--18]& 2.3056(28)(33)[11--18]\\
& $m_2$& 2.3104(27)(0)[7--13]& 2.3205(23)(23)[9--17]& 2.3173(33)(3)[8--18]& 2.3142(28)(33)[11--18]\\
\hline
$\O_{c}$& $m_1$& 1.7216(24)(1)[9--15]& 1.7240(24)(5)[12--18]& 1.7101(52)(77)[12--16]& 1.7160(39)(13)[12--18]\\
& $m_2$& 1.7261(24)(1)[9--15]& 1.7285(24)(5)[12--18]& 1.7146(52)(76)[12--16]& 1.7205(39)(13)[12--18]\\
\hline
$\Xi^\prime_{c}$& $m_1$& 1.6754(26)(32)[6--18]& 1.6799(29)(43)[9--16]& 1.6875(52)(57)[9--16]& 1.6881(43)(2)[11--18]\\
& $m_2$& 1.6799(26)(32)[6--18]& 1.6844(29)(43)[9--16]& 1.6920(52)(58)[9--16]& 1.6927(43)(2)[11--18]\\
\hline
$\Xi_{c}$& $m_1$& 1.6076(82)(86)[12--18]& 1.6078(48)(54)[12--18]& 1.6167(40)(9)[8--18]& 1.6120(41)(47)[12--17]\\
& $m_2$& 1.6121(82)(87)[12--18]& 1.6123(48)(55)[12--18]& 1.6211(40)(9)[8--18]& 1.6163(41)(48)[12--17]\\
\hline
$\S_{c}$& $m_1$& 1.6157(50)(38)[7--17]& 1.6252(55(0))[9--15]& 1.6446(56)(0)[8--16]& 1.6661(43)(70)[10--18]\\
& $m_2$& 1.6203(50)(38)[7--17]& 1.6298(55)(0)[9--15]& 1.6491(56)(0)[8--16]& 1.6706(43)(69)[10--18]\\
\hline
$\L_{c}$& $m_1$& 1.4974(71)(47)[6--13]& 1.523(16)(3)[12--18]& 1.5571(55)(22)[8--18]& 1.572(5)(18)[12--17]\\
& $m_2$& 1.5018(71)(48)[6--13]& 1.527(16)(3)[12--18]& 1.5615(55)(22)[8--18]& 1.577(5)(18)[12--17]\\
\end{tabular}
\end{ruledtabular}
\caption{\label{tab:baryons}{Charmed baryon masses in lattice units with 2 values of $m_0$ (indicated as $m_1 = 0.2034$ and $m_2 = 0.2100$) in Eq.~\eqref{eq:action_S_0}. The first uncertainty is statistical and the second is systematic from the different choice of fitting ranges (presented in square brackets).}}
\end{table}
%%%%%%%%%%%%%%%%
\begin{table}[tb]
\begin{ruledtabular}
\begin{tabular}{c|cllll}
Hadron& $m$& \texttt{m007}& \texttt{m010} &\texttt{m020}& \texttt{m030}\\
\hline
$\eta_{c}$
&$m_1$& 1.8783(4)(0)[14--19]& 1.8804(3)(0)[12--19]& 1.8687(4)(1)[12--19]& 1.8598(3)(2)[8--15] \\
& $m_2$& 1.8866(4)(1)[14--19]& 1.8887(3)(1)[12--19]& 1.8771(4)(1)[12--19]& 1.8683(5)(0)[8--15] \\
\hline
$J/\Psi$
&$m_1$& 1.9390(7)(0)[14--18]& 1.9421(4)(0)[10--19]& 1.9296(6)(1)[12--19]& 1.9198(6)(2)[11--19] \\
& $m_2$& 1.9470(7)(0)[14--18]& 1.9501(4)(1)[10--19]& 1.9376(6)(1)[12--19]& 1.9278(6)(3)[11--19] \\
\hline
$\chi_{c0}$
& $m_1$& 2.1660(54)(21)[9--16]& 2.1803(33)(6)[6--17]& 2.1652(55)(50)[6--18]& 2.1626(54)(2)[6--18] \\
& $m_2$& 2.1741(54)(20)[9--16]& 2.1883(35)(6)[6--17]& 2.1733(55)(49)[6--18]& 2.1705(54)(2)[6--18] \\
\hline
$\chi_{c1}$
&$m_1$& 2.2092(69)(24)[9--18]& 2.2234(52)(35)[9--16]& 2.2123(40)(8)[4--17]& 2.2004(44)(25)[4--17] \\
& $m_2$& 2.2171(69)(24)[9--18]& 2.2312(52)(35)[9--16]& 2.2199(40)(9)[4--17]& 2.2081(44)(25)[4--17] \\
\hline
$\chi_{c2}$
&$m_1$& 2.2224(64)(86)[6--18]& 2.2386(32)(24)[4--18]& 2.2205(45)(21)[4--17]& 2.2151(63)(26)[5--18] \\
& $m_2$& 2.2301(65)(85)[6--18]& 2.2463(32)(25)[4--18]& 2.2282(46)(19)[4--17]& 2.2226(63)(25)[5--18] \\
\end{tabular}
\end{ruledtabular}
\caption{\label{tab:mesons}{Charmonium masses in lattice units with $m_1 = 0.2034$ and $m_2 = 0.2100$.}}
\end{table}
%%%%%%%%%%%%%%%%

%%%%%%%%%%%%%%%%
%	FIGURE: EFF MASS PLOTS
%%%%%%%%%%%%%%%%
%%%%%%%%%%%%%%%%

%%%%%%%%%%%%
%
%		LIGHT- AND HEAVY-QUARK MASS EXTRAPOLATION
%
%%%%%%%%%%%%
\section{Heavy- and Light-Quark Mass Extrapolation\label{sec:ChPT}}
In order to make contact with experiment, we must extrapolate our results to infinite volume, continuum limit and to the physical value of the light- and heavy-quark masses.  Optimally, the extrapolations can be performed in terms of dimensionless ratios of observable quantities, so as to minimize contamination from a particular scale-setting method.  In this work, we have chosen to scale our masses by the calculated value of the pion decay constant on each ensemble, forming the dimensionless ratios $M_h / f_\pi$, where $M_h$ is the mass of a given hadron. We take the values of $f_\pi$ (and $m_\pi$) from Ref.~\cite{WalkerLoud:2008bp}; they are collected in Table~\ref{tab:mpifpi}.
%%%%%%%%%%%%%%%
%% TABLE mpi fpi
%%%%%%%%%%%%%%%
\begin{table}
\begin{ruledtabular}
\begin{tabular}{c|cccc}
ensemble: $\begin{array}{c}\beta \\ am_l \end{array}$
&$\begin{array}{c}6.76\\ 0.007\end{array}$
&$\begin{array}{c}6.76\\ 0.010\end{array}$
&$\begin{array}{c}6.79\\ 0.020\end{array}$
&$\begin{array}{c}6.81\\ 0.030\end{array}$ \\
%$am_l^{dwf}$ &0.0081& 0.0138& 0.0313& 0.0478 \\
\hline
$am_\pi$& 0.1842& 0.2238& 0.3113& 0.3752\\
$af_\pi$& 0.0929& 0.0963& 0.1026& 0.1076 \\
$m_\pi / f_\pi$& 1.983& 2.325& 3.035& 3.489
\end{tabular}
\end{ruledtabular}
\caption{\label{tab:mpifpi}Values of $m_\pi$ and $f_\pi$ calculated in Ref.~\cite{WalkerLoud:2008bp}. For all ensembles the staggered strange-quark mass is $am_s = 0.050$ while the domain-wall strange-quark mass is $am_s^{\rm dwf} = 0.081$.}
\end{table}
%%%%%%%%%%%%%%%%
As can be seen, $af_\pi$ varies by $\approx 15\%$ over the range of pion masses used in this work, adding additional chiral curvature. However, the light-quark mass dependence of $f_\pi$ is well understood~\cite{Gasser:1983yg,Colangelo:2005gd}, and so this variation can be accounted for.

Ultimately, one would like to use heavy-hadron chiral perturbation theory (HH$\chi$PT)~\cite{Wise:1992hn,Burdman:1992gh,Yan:1992gz,Cho:1992gg,Cho:1992cf,Savage:1995dw,Hu:2005gf} to perform both the charm-quark mass extrapolation and the chiral extrapolation of the charmed hadron masses, allowing a lattice determination of not just the spectrum but also the low-energy constants entering the effective field theory. There are several reasons we cannot perform a thorough extrapolation in this manner. First, we only have results at four independent values of the light-quark mass, and at only one value of the strange mass. Second, in this work, we only have results for the $J=1/2$ baryons, and a proper chiral extrapolation requires also the spectrum of $J=3/2$ charmed baryons; the states are related by the heavy-quark symmetry, and therefore the mass splittings are small (similarly, the extrapolation of the heavy meson masses requires the $J=1$ states as well as $J=0$). Third, our calculation is mixed-action, thus requiring either a continuum extrapolation or the use of mixed-action $\chi$PT~\cite{Bar:2005tu,Tiburzi:2005is,Chen:2006wf,Orginos:2007tw,Chen:2007ug,Chen:2009su}. The mixed-action effective field theory can be trivially constructed from the partially quenched theories for heavy hadrons~\cite{Savage:2001jw,Tiburzi:2004kd,Mehen:2006vv} by following the prescription in Ref.~\cite{Chen:2007ug}. However, this work only utilizes one lattice spacing, and so one can not perform the full mixed-action analysis. With these caveats in mind, we proceed with our analysis.

%%%%%%%%%%%%
%
%		FPI SCALE SETTING
%
%%%%%%%%%%%%
\subsection{Scale setting with $f_\pi$\label{sec:fpiScale}}
The light-quark mass expansion of a heavy-hadron mass is given by%
%FOOTNOTE
\footnote{Here we are presenting an $SU(2)$ extrapolation formula with the operator normalization of Ref.~\cite{Tiburzi:2008bk} such that the coefficient $c_h^{(2)}$ is dimensionless.}
\begin{equation}
M_h = M_0
	+\frac{c_h^{(2)}}{4\pi} \frac{2B m_l}{f_0} +\cdots
\end{equation}
At this order, we are free to make the replacements $f_0 \rightarrow f_\pi$ and $2Bm_l \rightarrow m_\pi^2$, with corrections appearing at $\mc{O}(m_\pi^4)$.  The dots represent terms of higher order in the chiral expansion, with the first non-analytic (in the quark mass) corrections appearing as corrections which scale as $\sim m_\pi^3$. As stated above, we are scaling our masses with $f_\pi$ to form dimensionless ratios for extrapolation,
\begin{equation}\label{eq:Mhfpi}
	\frac{M_h}{f_\pi} = \frac{M_0}{f_\pi} + \frac{c_h^{(2)}}{4\pi} \frac{m_\pi^2}{f_\pi^2} +  \cdots
\end{equation}
When performing an extrapolation in this manner, it is important to realize we cannot approximate $M_0 / f_\pi$ as a constant, since the chiral corrections to $f_\pi$ are $\mc{O}(m_\pi^2)$ and thus are the same order as the term with coefficient $c_h^{(2)}$. Rather, the chiral expansion of $f_\pi$ is given by~\cite{Gasser:1983yg} (with the normalization $f_0 \sim 130$~{MeV})
\begin{align}
f_\pi &= f_0 \left[ 1
	-\frac{2m_\pi^2}{(4\pi f_\pi)^2} \ln \left( \frac{m_\pi^2}{\mu^2} \right)
	+2 l_4(\mu) \frac{m_\pi^2}{f_\pi^2} \right]
	+\cdots
\nonumber\\&
	\equiv f_0 \Big[ 1 + \delta f(m_\pi/f_\pi) \Big] +\cdots
\end{align}
In this expression, we have made use of perturbation theory to replace all terms appearing at next-to-leading order with their (lattice) physical values.  Similarly, we have rescaled the renormalization scale $\mu \rightarrow  \tilde{\mu}\, f_\pi$ to express the chiral corrections as purely a function of $m_\pi / f_\pi$.  Again, the corrections to this rescaling first appear at next-to-next-to-leading order.
In order to perform our chiral extrapolations using Eq.~\eqref{eq:Mhfpi}, we must determine $l_4$, which captures the chiral corrections of $f_\pi$.  The mixed-action formula for $f_\pi$ is known~\cite{Bar:2005tu}, but again, only useful if one has data for at least two lattice spacings. Since we currently only have results at one lattice spacing, we perform a continuum chiral extrapolation analysis of the $af_\pi$ in Table~\ref{tab:mpifpi}. The results are collected in Table~\ref{tab:fpil4}.
%%%%%%%%%%%%%%%
%% TABLE fpi FIT
%%%%%%%%%%%%%%%
\begin{table}
\begin{ruledtabular}
\begin{tabular}{c|ccc}
range& \texttt{m007}--\texttt{m010}& \texttt{m007}--\texttt{m020}& \texttt{m007}--\texttt{m030}\\
\hline
$l_4(\mu = f_\pi)$& 0.0307(27)& 0.0293(6)& 0.0302(4)\\
\end{tabular}
\end{ruledtabular}
\caption{\label{tab:fpil4}Values of $l_4$ needed for chiral extrapolations of $M_h / f_\pi$.  The different values of $l_4$ are determined through the different choices of fitting range, also listed.}
\end{table}
%%%%%%%%%%%%%%%%
%%%%%%%%%%%%%%%%
%%%  FIGURE: f_pi
\begin{figure}[tb]
\begin{tabular}{cc}
\includegraphics[width=0.5\textwidth]{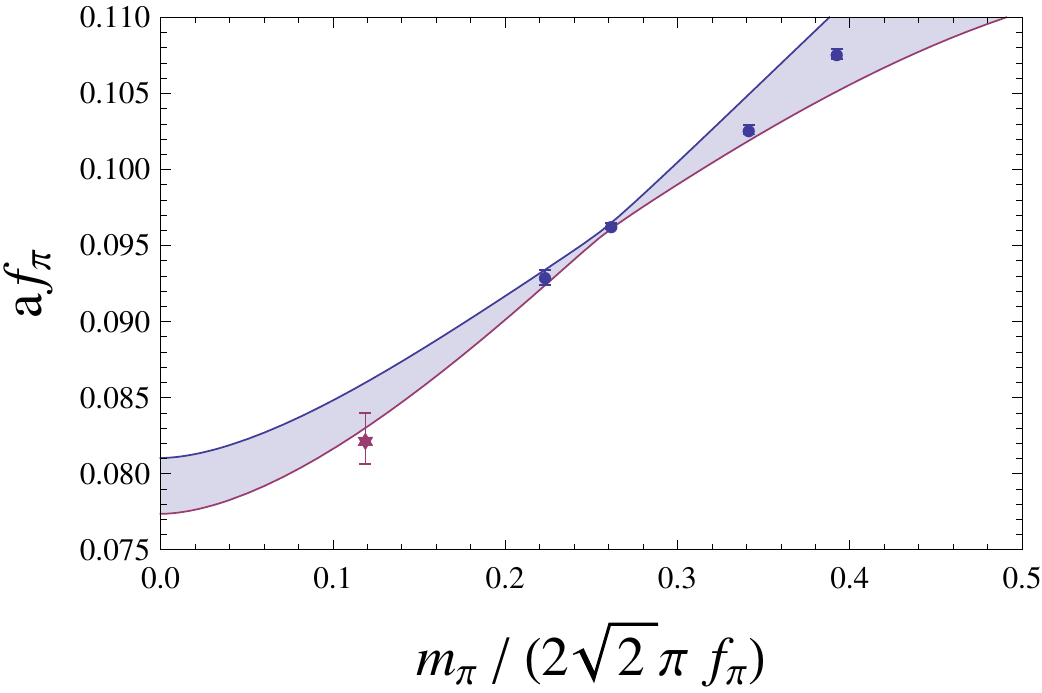}
&
\includegraphics[width=0.5\textwidth]{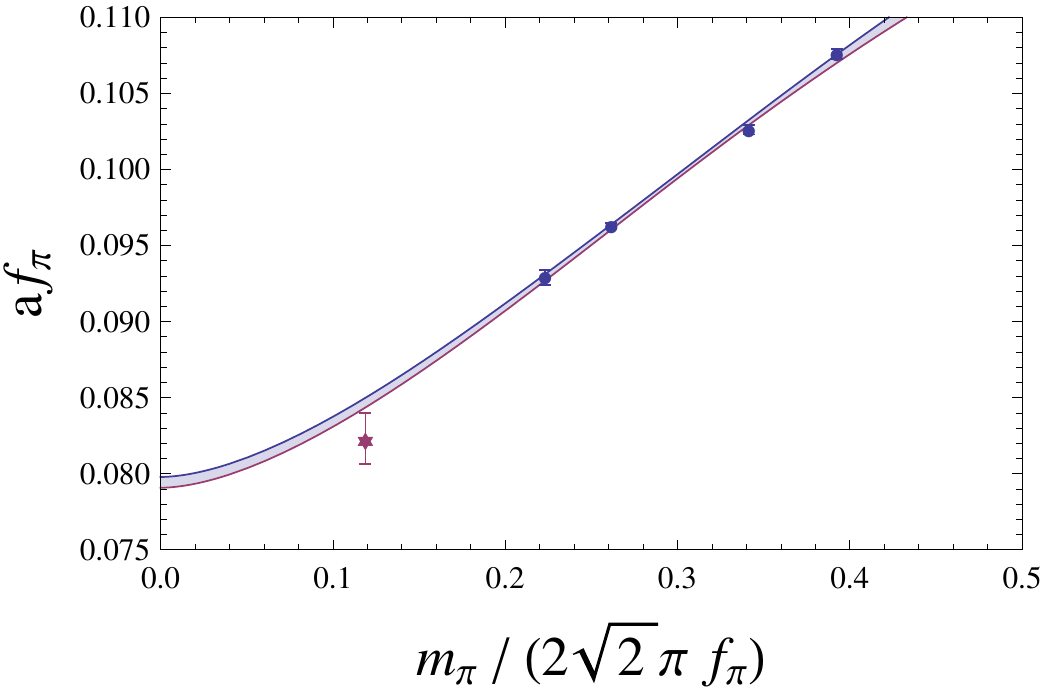}
\\ (a) & (b)
\end{tabular}
\caption{\label{fig:fpiExtrap}The (blue) filled circles represent the lattice data and the (red) star is the physical point, converted to lattice units using $a^{-1} = 1588$~{MeV} with a 2\% error bar added for the scale setting. The error bands are the 68\% confidence intervals in the resulting chiral extrapolation from the lightest two points (a) and a fit to all four lattice points (b).}
\end{figure}
%%%%%%%%%%%%%%%%
%%%%%%%%%%%%%%

The resulting extrapolations are plotted in Figure~\ref{fig:fpiExtrap}.  In this figure, the (blue) filled circles are the lattice data, and the error bands represent the 68\% confidence intervals.  The (red) star denotes the physical value converted to lattice units using $a^{-1} = 1588$~{MeV}~\cite{Aubin:2004fs}. We assign an additional 2\% error to this point to estimate the uncertainty in the scale setting method.  In Figure~\ref{fig:fpiExtrap}(a) we display the fit to the lightest two points and in (b) the fit to all four points.  Note that the extrapolation describes the values of $f_\pi$ very well.  Additionally, one sees that using $f_\pi$ or $r_1$ to set the scale results in agreement in the extrapolated values, as first observed in Ref.~\cite{Beane:2005rj}.\footnote{The scale of $r_1$ is determined through the static-quark potential by solving for $r_1^2 F(r_1) = 1$; the values of $r_1/a$ can be found in Ref.~\cite{Bazavov:2009bb}.}
%

%%%%%%%%%%%%
%
%		HEAVY-QUARK MASS EXTRAPOLATION
%
%%%%%%%%%%%%
\subsection{Charm-Quark Mass Extrapolation\label{sec:HQactionTest}}
\begin{figure}
\begin{tabular}{cc}
\includegraphics[width=0.5\textwidth]{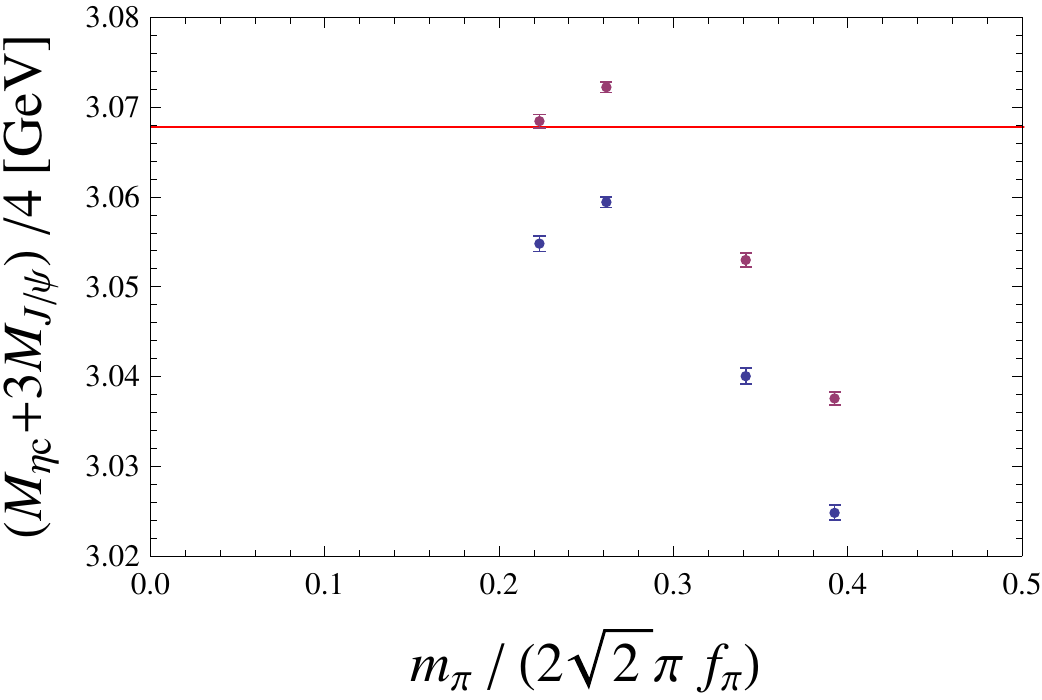}
&
\includegraphics[width=0.5\textwidth]{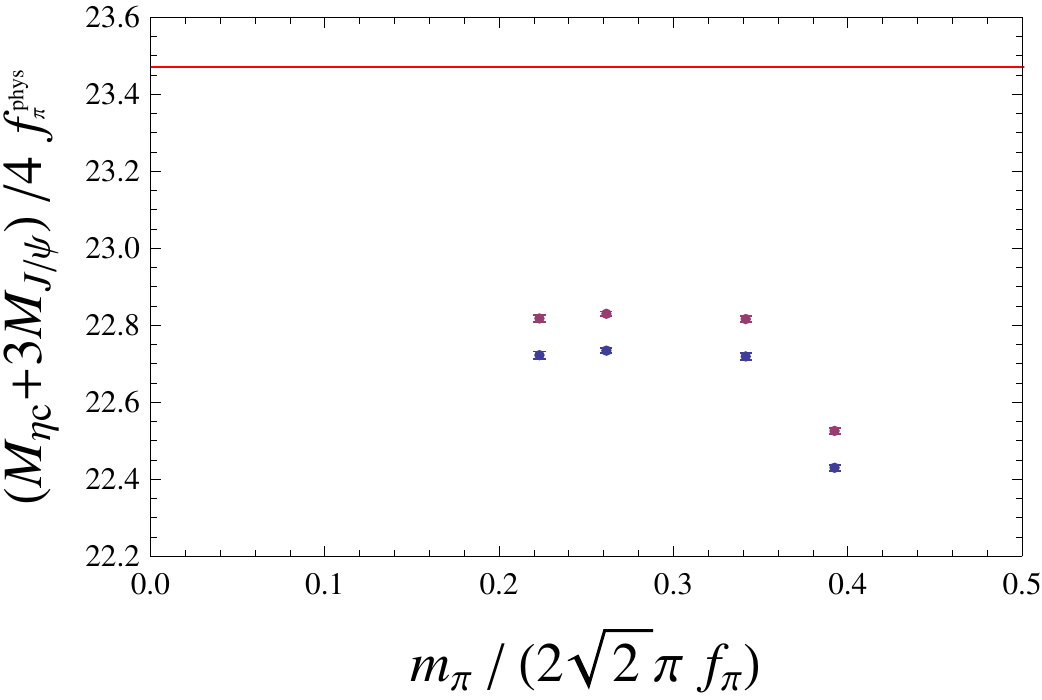}
\end{tabular}
\caption{\label{fig:spin_average_mass}Spin-averaged mass of $\eta_c$ and $J/\Psi$ on the different ensembles. The blue points and purple points indicate the masses at $m_1$ and $m_2$ respectively. The red line indicates the experimental value.  The left panel displays the results from the lattice spacing $a^{-1} = 1588~{\rm MeV}$ used on all ensembles.  This method was used to tune the charm-quark mass on the \texttt{m007} ensemble.  The right panel displays the masses scaled by $f_\pi$ on the lattice and extrapolated to $f_\pi^{\rm phys}$, as discussed in the text.}
\end{figure}

To tune the charm-quark mass we use the spin-averaged $J/\Psi$-$\eta_c$ mass.  We use the lattice spacing determined by MILC ($a^{-1} =1588$~MeV~\cite{Aubin:2004fs}) on the \texttt{m007} ensemble to estimate the two charm-quark masses used for our charm quark propagator calculations.  These same two charm quark masses, $m_1$ and $m_2$, were used on all ensembles.  On the MILC ensembles, the value of $\beta$ was slightly varied for the different light-quark masses.  Therefore, the corresponding value of the critical mass changes from ensemble to ensemble, leading to a slightly different charm-quark mass tuning.  This can be clearly seen in the left panel of Fig.~\ref{fig:spin_average_mass}, where we display the spin-averaged $J/\Psi$--$\eta_c$ mass as a function of the light-quark mass, determined with the $a^{-1} = 1588$~MeV scale setting.  Ensembles \texttt{m007} and \texttt{m010} share the same value of $\beta$ and therefore the difference in these points (the left-most two sets of masses) is due entirely to light-quark contributions, whereas the \texttt{m020} and \texttt{m030} ensembles each have a different value of $\beta$, so that the variation of the spin-averaged mass is due both to light-quark effects as well as a shifted value of the critical mass.
%%%%%%%%%%%%%%%%%%
%%	FIGURE: Spin average mass of J/\Psi and \eta_c
%%%%%%%%%%%%%%%%%%
%%%%%%%%%%%%%%%%%%

In the right panel of Fig.~\ref{fig:spin_average_mass}, we display our preferred method of determining the charm-quark mass using $f_\pi$ to set the scale.  On each ensemble, we take the spin-averaged $J/\Psi$-$\eta_c$ mass and divide by the corresponding value of $f_\pi^{\rm latt}$ calculated on that ensemble.  We then use the value of $l_4$ determined in Sec.~\ref{sec:fpiScale} to scale these values to determine the ratio with $f_\pi^{\rm phys}$,
\begin{align}
\frac{M_{\eta_c} +3M_{J/\Psi}}{4f_\pi^{\rm  phys}} &=
	\frac{1+\d f(m_\pi^{\rm  latt} / f_\pi^{\rm  latt})}{1 + \d f(m_\pi^{\rm phys} / f_\pi^{\rm phys})}
	\frac{ M_{\eta_c} +3M_{J/\Psi}}{4f_\pi^{\rm latt}}\, .
\end{align}
It is these scaled values that are plotted in the right panel of Fig.~\ref{fig:spin_average_mass} and which we use to extrapolate our spectrum calculation to the physical charm-quark mass point, which we take to be
\begin{align}
&&	&\frac{M_{\eta_c}^{\rm phys} + 3 M_{J/\Psi}^{\rm phys}}{4 f_\pi^{\rm phys}} = 23.47\,,&\\
&\textrm{with}&
&\frac{m_\pi^{\rm phys}}{f_\pi^{\rm phys}} = 1.056\, .&
\end{align}
Here, $m_\pi^{\rm phys}$ is taken to be the isospin-averaged pion mass, while $f_\pi^{\rm phys}$ is taken to be the charged-pion decay constant~\cite{Amsler:2008zzb}.
On each ensemble, we linearly extrapolate the spin-averaged $J/\Psi$-$\eta_c$ mass (scaled by $f_\pi^{\rm phys}$) to the experimental value to determine the parameter $m_0=m_c^{\rm phys}$ (the masses of all  hadrons are then extrapolated linearly to this charm-quark mass  on each ensemble). The uncertainties of the extrapolated hadron masses  are evaluated using the jackknife method.  As a check of systematics, we perform the same procedure using the lattice spacing $a^{-1} =1588$~MeV to perform the linear charm-quark mass extrapolation. Using this second approach, the resulting charmed baryon spectrum is consistent with that of our preferred charm-quark mass-tuning method.

\begin{figure}
\begin{tabular}{ccc}
\includegraphics[width=0.33\textwidth]{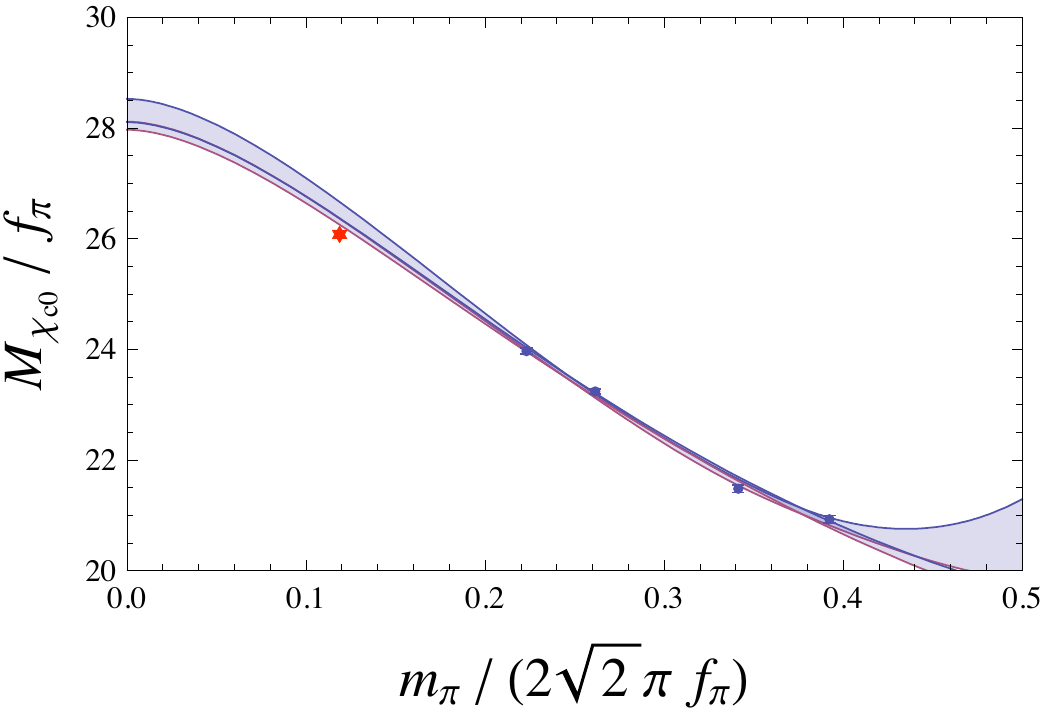}&
\includegraphics[width=0.33\textwidth]{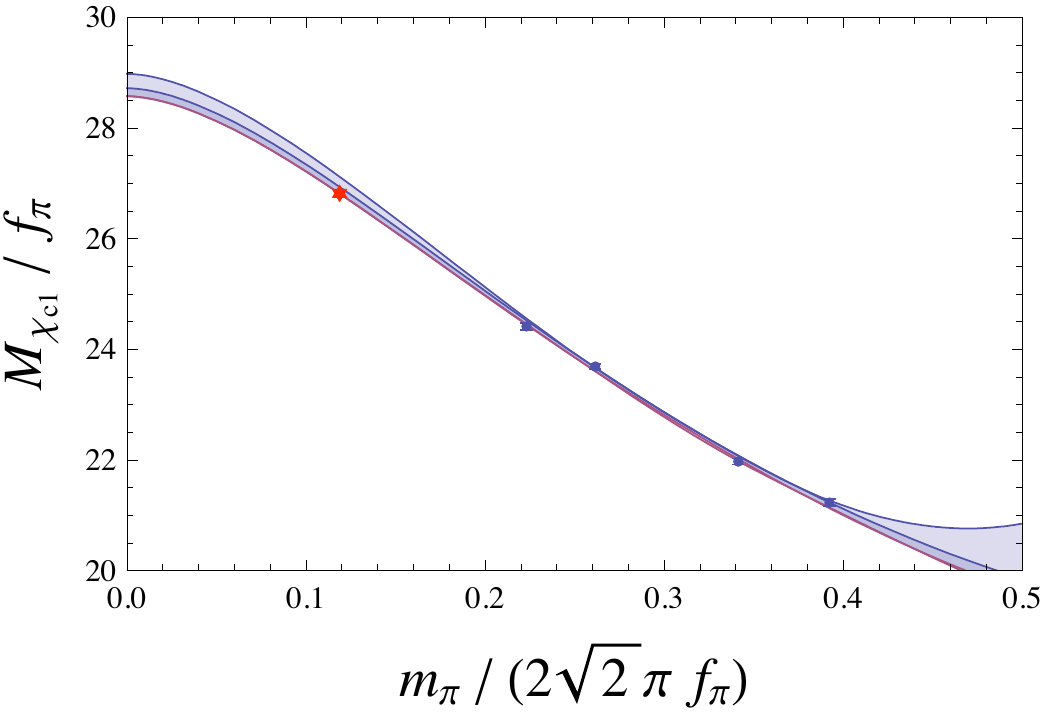}&
\includegraphics[width=0.33\textwidth]{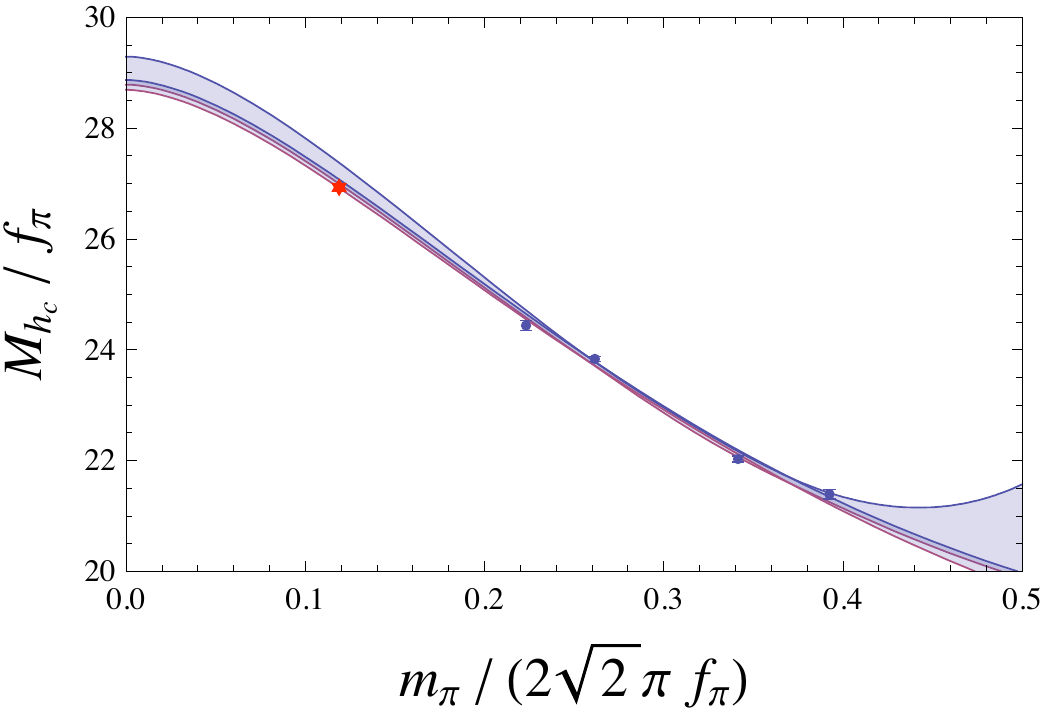}
\end{tabular}
\caption{\label{fig:charmonia_mass}The masses of $\chi_{c0}$, $\chi_{c1}$ and $h_{c}$ as functions of $m_\pi^2/f_\pi^2$. The blue points are our numerical values. The blue shaded regions show the standard deviation allowed regions of linear fit. The pink shaded regions show the standard deviation allowed regions of quadratic fit. The red points are experimental values.}
\end{figure}

To test the viability of our choice of mixed-action and to gauge the discretization errors, we compute both the $J/\Psi$-$\eta_c$ hyperfine mass splitting as well as the low-lying charmonium spectrum of the $\chi_{c0}$, $\chi_{c1}$ and $h_{c}$.  The interpolating fields used for these charmonium states are%
%FOOTNOTE
\footnote{One can also use improved interpolating operators to extract charmonium states in lattice calculations, especially for the excited states $\chi_{c0}$, $\chi_{c1}$ and $h_{c}$; see, for example, Ref.~\cite {Dudek:2007wv}. }
\begin{align}
&\chi_{c0} = \bar{Q}_c\, Q_c\, ,& \\
&\chi_{c1}^i = \bar{Q}_c\, \g^i \g_5 Q_c\, ,& \\
&h_{c}^{i} = \sum_{j=1}^{3}\sum_{k=j}^{3}\epsilon_{ijk}\bar{Q}_c\, \g^j \g^k Q_c\, ,&
\end{align}
To extrapolate these charmonium masses to the physical light-quark mass values, we use Eq.~\eqref{eq:Mhfpi} both in quadratic (in $m_\pi$) as well quartic form, i.e.
\begin{equation}\label{eq:MhfpiQuartic}
	\frac{M_h}{f_\pi} = \frac{M_0}{f_\pi} + \frac{c_h^{(2)}}{4\pi} \frac{m_\pi^2}{f_\pi^2} +   \frac{c_h^{(4)}}{(4\pi)^2} \frac{m_\pi^4}{f_\pi^4} \;.
\end{equation}
The results of the extrapolation are displayed in Fig.~\ref{fig:charmonia_mass}, and tabulated in Tab.~\ref{tab:chiC}.
%%%%%%%%%%%%%%%%%%
%%	FIGURE: Chi Masses
%%%%%%%%%%%%%%%%%%
%%%%%%%%%%%%%%%%%%
%%	Table: Chi Masses
%%%%%%%%%%%%%%%%%%
\begin{table}
\begin{ruledtabular}
\begin{tabular}{c|lll}
&$M_{\chi_{c0}}$ (MeV)& $M_{\chi_{c1}}$ (MeV)& $M_{h_{c}}$ (MeV) \\
\hline
Extrapolated Values& 3465(20)(13)& 3525(20)(6)& 3553(25)(14)\\
Experimental Values& 3415& 3511& 3526
\end{tabular}
\end{ruledtabular}
\caption{\label{tab:chiC}Low-lying charmonium spectrum of $\chi_{c0}$, $\chi_{c1}$ and $h_{c}$. The experimental values are taken from the Particle Data Group~\cite{Amsler:2008zzb}.}
\end{table}
%%%%%%%%%%%%%%%%%%
In the table, the first uncertainty is statistical and the second is an extrapolation systematic from the two extrapolation functions used.

\begin{figure}
\includegraphics[width=0.5\textwidth]{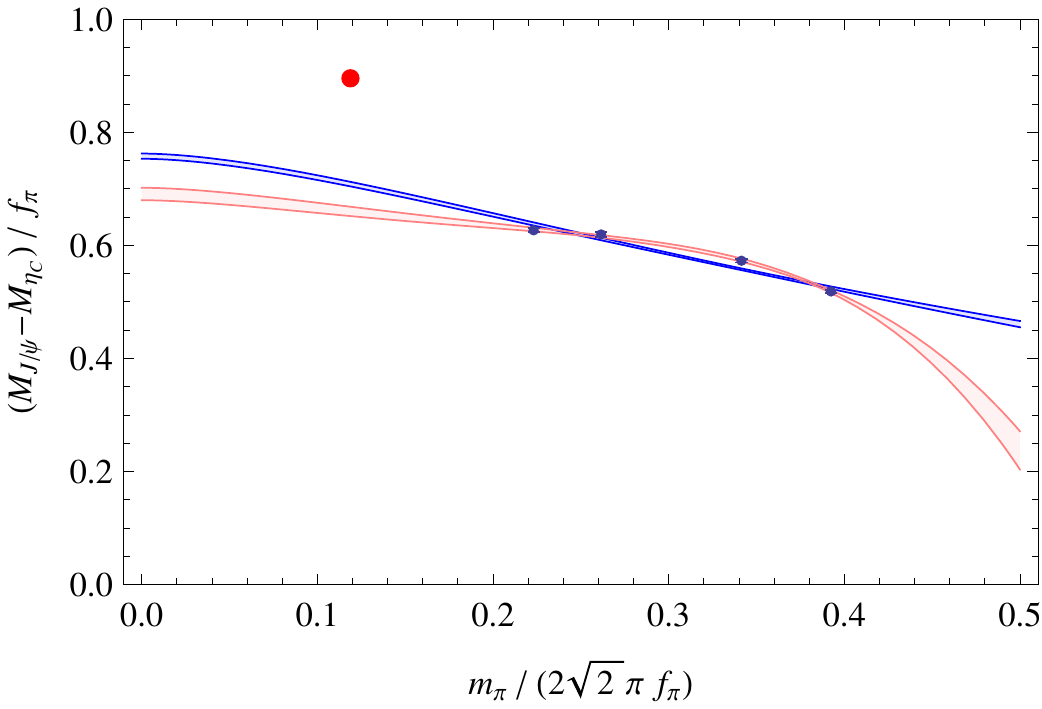}
\caption{\label{fig:hyp_splitting} {Extrapolation of the hyperfine splitting. The blue points are the lattice data. The red point is the experimental value. The blue band is the quadratic fit with Eq.~\ref{eq:Mhfpi}, while the pink band is the quartic fit with Eq.~\ref{eq:Mhfpi}.}}
\end{figure}

A more stringent test of discretization errors is the calculation of the hyperfine splitting. The hyperfine splitting is obtained by fitting the ratio of the two-point correlation functions of $J/\Psi$ and $\eta_c$
\begin{equation}
	\mathcal{R}=\frac{C_{J/\Psi}(t)}{C_{\eta_c}(t)}
\end{equation}
to a single exponential
\begin{equation}
	\mathcal{R}=Ae^{-\Delta_m t},
\end{equation}
where $\Delta_m$ is the mass splitting between the $J/\Psi$ and $\eta_c$. The splittings are first extrapolated to the physical charm-quark mass for each ensemble and then extrapolated to the physical light-quark mass.  As with the charmonium spectrum, we perform a light-quark mass extrapolation using both a quadratic and quartic form of Eq.~\eqref{eq:Mhfpi}.  In Fig.~\ref{fig:hyp_splitting} we display this extrapolation, finding $M_{J/\Psi} - M_{\eta_c} = 93(1)(7)$~MeV.  The first uncertainty is statistical while the second is a systematic from the chiral extrapolation.
%%%%%%%%%%%%%%%%%%%%%%
%%%%%%%%%%%%%%%%%%%%%%

It is well known that the lattice computations of the charmonium hyperfine splitting (experimentally measured to be 117~MeV) are sensitive to the lattice spacing.  Qualitatively, one can understand this by performing a Symanzik expansion of the heavy quark action, revealing dimension five operators arising from discretization effects, which are otherwise identical to the heavy quark effective theory (HQET)~\cite{Grinstein:1990mj,Georgi:1990um,Falk:1990yz} operator responsible for the hyperfine splitting%
%FOOTNOTE
\footnote{A proper treatment of heavy quark discretization effects is more involved and can be found in Ref.~\cite{ElKhadra:1996mp}.} 
\begin{align}
&\mc{L}_{HQET} \supset -g\, \bar{h}^{(+)}_c\, \frac{\mathbf{\s} \cdot \mathbf{B}}{2 m_c}\, h^{(+)}_c& \nonumber\\
	\longrightarrow\quad 
&\mc{L}_{\rm latt} \supset -g\, \bar{h}^{(+)}_c\,  \frac{\mathbf{\s} \cdot \mathbf{B}}{2 m_c}\, h^{(+)}_c
		+a\, c(am_c)\, \bar{h}^{(+)}_c\, \mathbf{\s}\cdot\mathbf{B}\, h^{(+)}_c\, ,&
\end{align}
where $h_c^{(+)}$ is the heavy quark field.  In the heavy quark action we are using, the coefficients of the operators $S_B$~\eqref{eq:action_S_B} and $S_E$~\eqref{eq:action_S_E} have been given their tree-level, tadpole improved values in order to mitigate the effects of this unwanted discretization effect.  It is known the operator $S_B$~\eqref{eq:action_S_B} has a significant effect on the hyperfine splitting~\cite{ElKhadra:1996mp,Christ:2006us,Lin:2006ur}.  A nonperturbative tuning of the coefficient $c_B$ can improve the hyperfine splitting in a fixed-lattice spacing calculation; see Ref.~\cite{Lin:2007qu}, in particular Fig.~3.  However, the qualitative aspects of this effect remain even after tuning the coefficients.  Previous quenched calculations of the hyperfine splitting have generally been low, being about 80~MeV, and showed a strong lattice-spacing dependence.
Further, a recent direct calculation of the disconnected diagrams has ruled out these (or their lack thereof) being the cause of the discrepancy~\cite{Levkova:2008qr}.
Our results are consistent with those of the Fermilab/MILC Collaboration, which utilized a similar heavy quark action, the same dynamical ensembles and staggered light quarks~\cite{Gottlieb:2005me}. The Fermilab/MILC Collaboration also performed calculations on different lattice spacings, finding similar lattice-spacing dependence to Ref.~\cite{Gottlieb:2005me}.
Therefore, the discrepancy of our calculated hyperfine splitting with the experimental value is expected.

%%%%%%%%%%%%
%
%		LIGHT-QUARK MASS EXTRAPOLATION
%
%%%%%%%%%%%%
\subsection{Light-Quark Mass Extrapolation}

%%%%%%%%%%%%
%
%		HHChPT
%
%%%%%%%%%%%%
\subsubsection{Heavy-Hadron $\chi$PT Extrapolation}
\begin{figure}[tb]
\begin{tabular}{cc}
\includegraphics[width=0.5\textwidth]{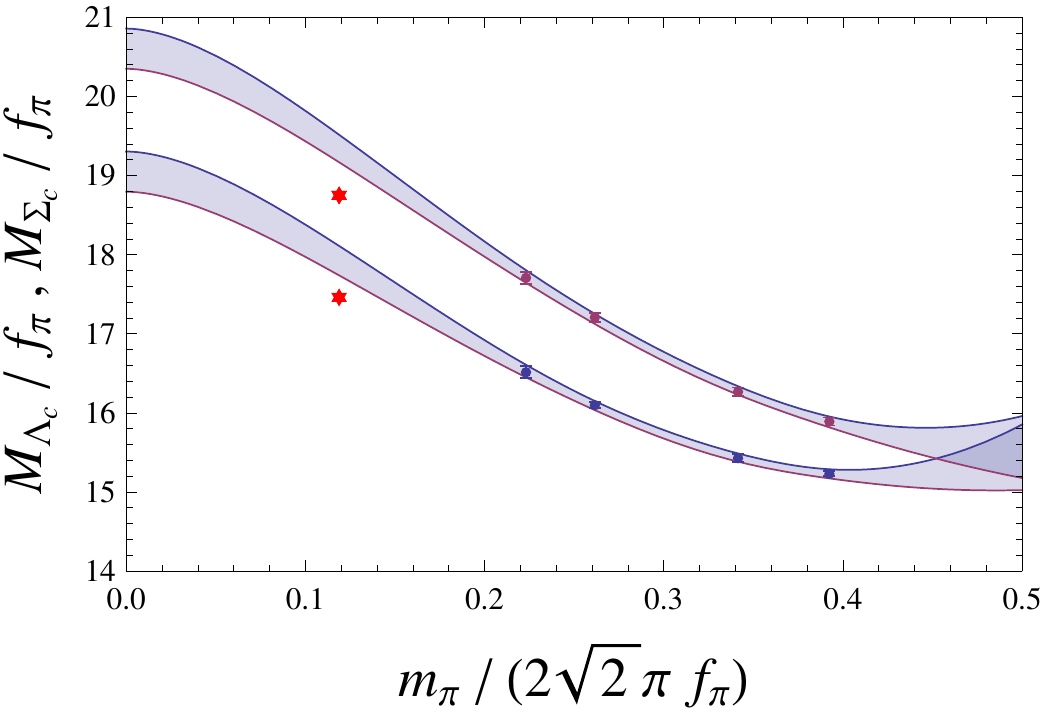}
&
\includegraphics[width=0.5\textwidth]{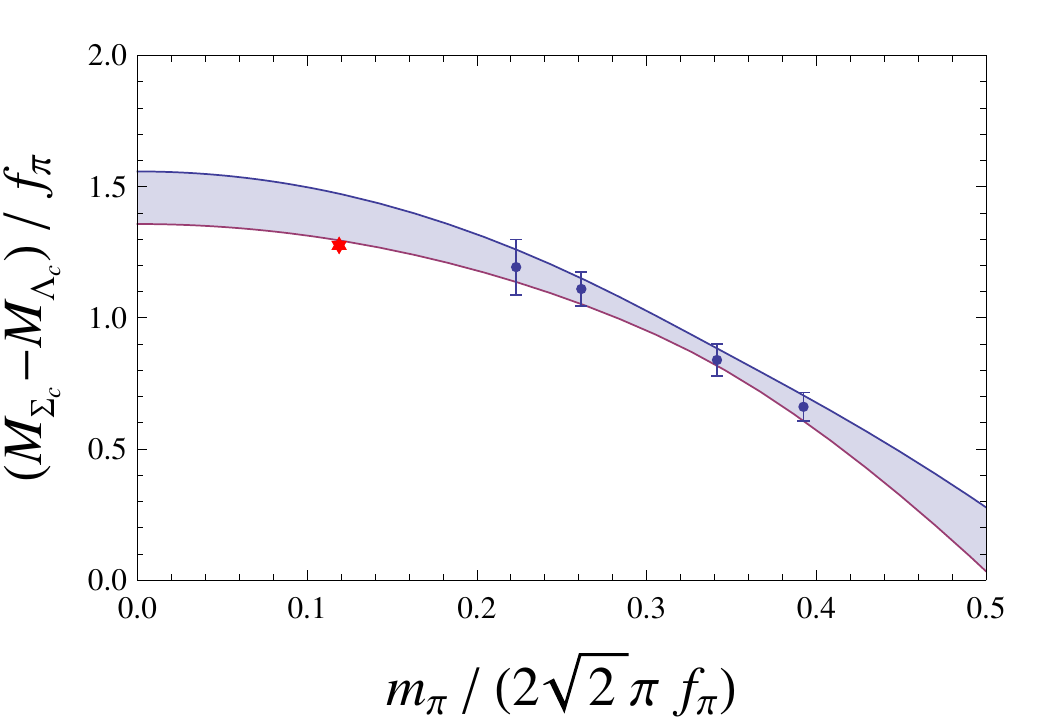}
\\
(a) & (b)
\end{tabular}
\caption{\label{fig:LamSigChPT}NLO HH$\chi$PT extrapolation of $M_{\L_c}$ and $M_{\S_c}$ (a) as well as $M_{\S_{c}}-M_{\L_c}$ (b).}
\end{figure}

To perform the light-quark mass extrapolation, we begin with a continuum HH$\chi$PT extrapolation of the baryon masses. The mass formula for these baryons containing a heavy quark was first determined in Ref.~\cite{Savage:1995dw} and later extended to partially quenched theories in Ref.~\cite{Tiburzi:2004kd}. For doubly heavy baryons, the $\chi$PT was formulated in Ref.~\cite{Hu:2005gf} and later extended to partially quenched theories in Ref.~\cite{Mehen:2006vv}.  In this work, we perform $SU(2)$ chiral extrapolations of the baryon masses, inspired by Ref.~\cite{Tiburzi:2008bk}.%
%FOOTNOTE
\footnote{For further discussion on $SU(2)$ chiral extrapolations of hadron states with strange valence quarks, see Refs.~\cite{Jiang:2009sf,Mai:2009ce,Tiburzi:2009cf}.}
To perform the extrapolations, we treat the $J=1/2$ and $J=3/2$ baryons as degenerate, which is valid at this order in HQET/HH$\chi$PT.%
%FOOTNOTE
\footnote{It would be more desirable to use the lattice-calculated masses of the $J=3/2$ baryons, but we do not have them for this work, and so we use this approximation for now.}
The baryons are grouped into their respective $SU(2)$ multiplets allowing for a simultaneous two-flavor chiral extrapolation of all masses in related multiplets. This allows us, with only four gauge ensembles, to determine all the relevant LECs for a given pair of multiplets in a global fit. The first pair of multiplets contains the $\L_c$ and $\S_c$ baryons. Their $SU(2)$ chiral extrapolation functions are given at next-to-leading order (NLO) by
\begin{align}\label{eq:MLamC}
\frac{M_{\L_c}}{f_\pi} &= \frac{M_0}{f_0}\frac{1}{1 + \d f(m_\pi/f_\pi)}
	- \frac{c_\L^r(\mu)}{4\pi}\frac{m_\pi^2}{f_\pi^2}%\Big[1 + \d f(m_\pi,f_\pi,\mu) \Big]^n
	-\frac{6 g_{3}^2}{(4\pi)^2} \frac{\mc{F}(m_\pi,\D_{\S\L},\mu)}{f_\pi^3}\, ,
\\ \nonumber\\
\label{eq:MSigC}
\frac{M_{\S_c}}{f_\pi} &= \frac{M_0+ \D_{\S\L}^{(0)}}{f_0}\frac{1}{1 + \d f(m_\pi/f_\pi)}
	- \frac{c_\S^r(\mu)}{4\pi}\frac{m_\pi^2}{f_\pi^2}%\Big[1 + \d f(m_\pi,f_\pi,\mu) \Big]^n
\nonumber\\&\qquad	
	-\frac{2}{3}\frac{g_{3}^2}{(4\pi)^2} \frac{\mc{F}(m_\pi,-\D_{\S\L},\mu)}{f_\pi^3}
	+\frac{4}{3}\frac{g_{2}^2}{(4\pi)^2} \frac{\mc{F}(m_\pi,0,\mu)}{f_\pi^3}\, .
\end{align}
%where we vary $n = 0,1,2$ as an estimate of higher order corrections.
The chiral functions are
\begin{equation}
\mc{F}(m,\D,\mu) =
	(\D^2 - m^2 +i\e)^{3/2}
	\ln \left( \frac{\D+\sqrt{\D^2 - m^2 + i\e}}{\D - \sqrt{\D^2 - m^2 + i\e}} \right)
	-\frac{3}{2}\D m^2 \ln \left( \frac{m^2}{\mu^2}\right)
	-\D^3 \ln \left(\frac{4\D^2}{m^2}\right)\, .
\end{equation}
with
\begin{equation}
	\mc{F}(m,0,\mu) = \pi m^3\, ,
\end{equation}
and
\begin{align}
\mc{F}(m,-\D,\L)&= \left\{ \begin{array}{lc}
	-\mc{F}(m,\D,\L) +2i\pi (\D^2 - m^2)^{3/2}, & m < |\D| \\
	-\mc{F}(m,\D,\L) +2\pi (m^2 - \D^2)^{3/2}, & m > |\D|
	\end{array} \right. \, .
\end{align}
To stabilize the fits, we first fit $M_{\S_c} - M_{\L_c}$ to a quadratic in $m_\pi/f_\pi$,  and feed this into a fit of the masses, yielding the results in Table~\ref{tab:LamSigChPT} and extrapolations displayed in Figure~\ref{fig:LamSigChPT}.
One observes that the continuum HH$\chi$PT fits describe the lattice data very well.  However, only the leading term, $M_0$ is well determined,%
%FOOTNONTE
\footnote{To determine $M_0 / f_\pi^{\rm phys}$ we take our results for $M_0 / f_0$ and scale them by $[1+\d f(m_\pi^{\rm phys}/f_\pi^{\rm phys})]^{-1}$.}
while the rest of the LECs, most notably the axial couplings, $g_{\S\S\pi}$ and $g_{\S\L\pi}$ are consistent with zero. This phenomenon is not unique to the charmed baryons. In Ref.~\cite{WalkerLoud:2008bp}, chiral extrapolations on the nucleon mass in which the nucleon axial coupling, $g_{\pi NN}$ (commonly denoted as $g_A$ in baryon $\chi$PT) was left as a free parameter, returned values which were inconsistent with experiment and phenomenology.  In fact, given the lattice results for the nucleon mass as a function of $m_\pi$, it was found that the nucleon scales linearly in $m_\pi$. Such behavior signals a delicate cancelation between different orders, a trend which is found in all $2+1$ dynamical lattice computations of the nucleon mass~\cite{WalkerLoud:2008pj}.  Therefore, our findings for the axial couplings of the charmed baryons are not surprising in this light. To improve the situation, a simultaneous fit of the axial charges themselves, along with the masses will most likely be necessary.
%%%%%%%%%%%%%%%%
%%%  Lam Sig Fits
%%%%%%%%%%%%%%%%
\begin{table}[tb]
\begin{ruledtabular}
\begin{tabular}{c|c|ccccc|ccc}
Fit Range& $\D_{\S\L}/f_\pi^{\rm phys}$& $M_0 / f_\pi^{\rm phys}$& $c_\L^r(f_\pi)$& $c_\S^r(f_\pi)$& $g_{2}^2$& $g_{3}^2$& $\chi^2$& dof& $Q$\\
\hline
\texttt{m007}--\texttt{m030}&
	1.46(10)& 17.9(2)& -0.8(5)& 0.2(1.2)& 0.8(1.0)& $-$0.1(1)& 0.32& 3& 0.95\\
%\texttt{m007}--\texttt{m030}& 1&
%	1.46(10)& 17.9(2)& -0.7(4)& 0.2(1.3)& 0.9(1.5)& -0.08(12)& 0.33& 3& 0.96\\
\end{tabular}
\end{ruledtabular}
\caption{\label{tab:LamSigChPT}{Fit to $\L_c$ and $\S_c$ masses with NLO continuum formulae.}}
\end{table}
%%%%%%%%%%%%%%%%
%%%%%%%%%%%%%%%%

\begin{figure}[tb]
\begin{tabular}{cc}
\includegraphics[width=0.5\textwidth]{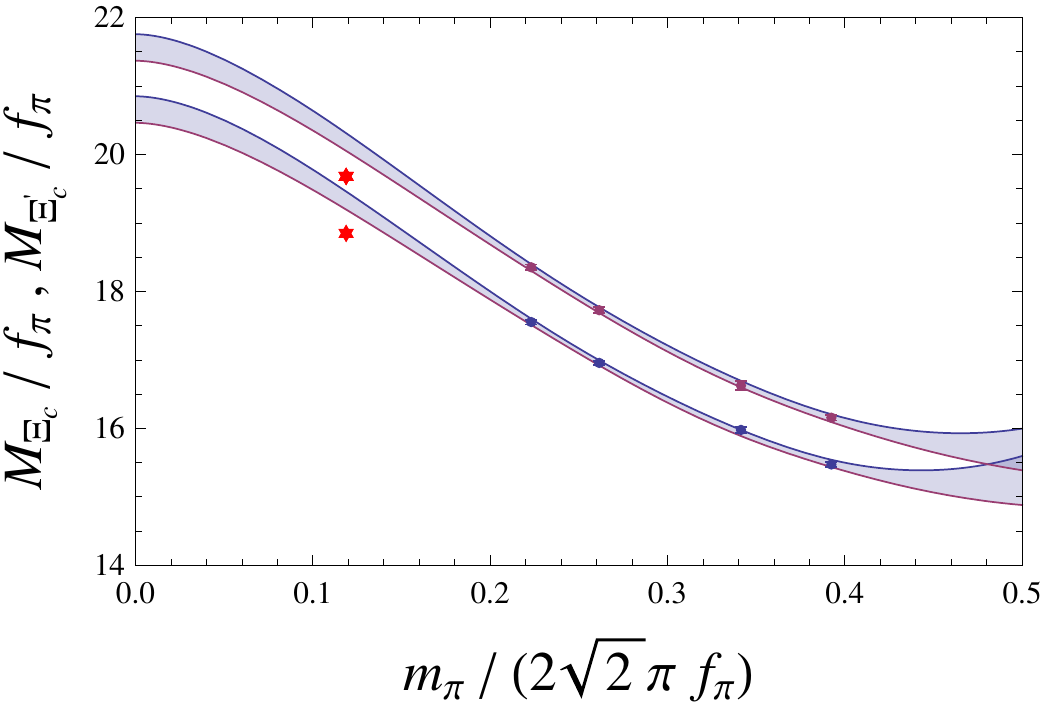}
&
\includegraphics[width=0.5\textwidth]{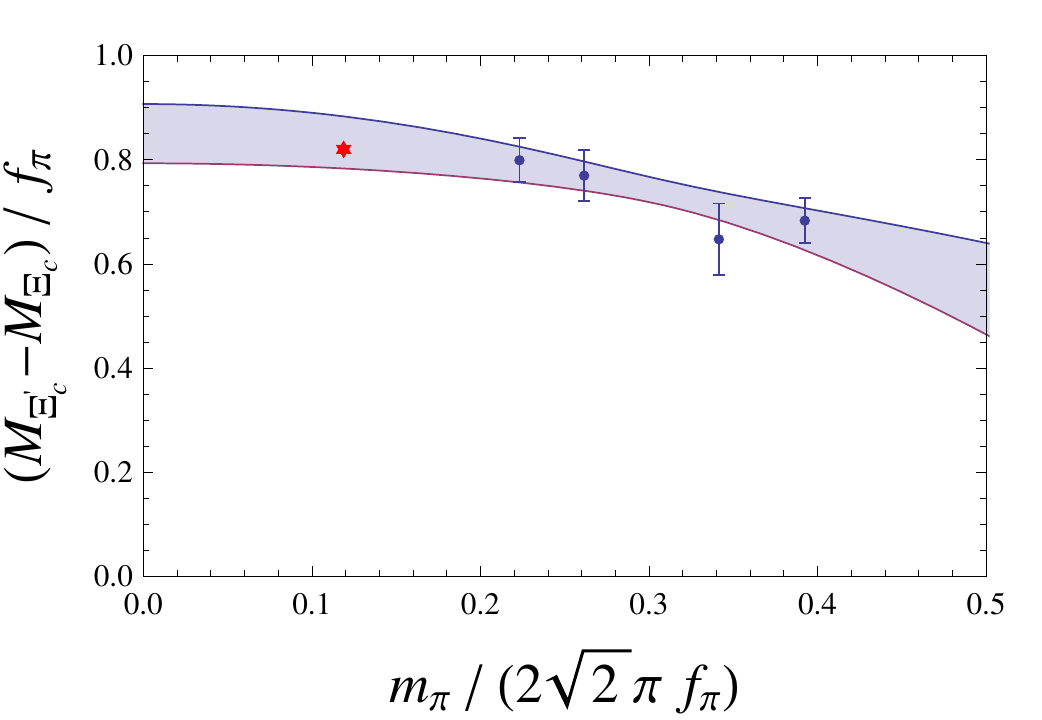}
\\
(a) & (b)
\end{tabular}
\caption{\label{fig:XiXip}NLO HH$\chi$PT extrapolation of $M_{\Xi_c}$ and $M_{\Xi^\prime_c}$ (a) as well as $M_{\Xi^\prime_{c}}-M_{\Xi_c}$ (b).}
\end{figure}

We perform a similar analysis for the $J=1/2$ $\Xi_c$-$\Xi_c^\prime$ isospin doublets, the results of which are collected in Table~\ref{tab:XiXipChPT} and displayed in Figure~\ref{fig:XiXip}. The extrapolation formulae for $M_{\Xi^\prime_c}$ and $M_{\Xi_c}$ are similar to those for $M_{\S_c}$ and $M_{\L_c}$. They can be deduced by comparing Eqs.~\eqref{eq:MLamC} and \eqref{eq:MSigC} to Ref.~\cite{Tiburzi:2004kd},%
%FOOTNOTE
\footnote{In $SU(3)$ HH$\chi$PT, the axial couplings for the $\Xi_c$-$\Xi_c^\prime$ system are the same as those for the $\L_c$-$\S_c$ system, $g_2 = g_{\S\S\pi} = g_{\Xi^\prime\Xi^\prime\pi}$ and $g_3 = g_{\S\L\pi} = g_{\Xi^\prime\Xi\pi}$. However, in the $SU(2)$ theories, they differ by $SU(3)$ breaking corrections.}
\begin{align}\label{eq:MXiC}
\frac{M_{\Xi_c}}{f_\pi} &= \frac{M_0}{f_0}\frac{1}{1 + \d f(m_\pi/f_\pi)}
	- \frac{c_\Xi^r(\mu)}{4\pi}\frac{m_\pi^2}{f_\pi^2}%\Big[1 + \d f(m_\pi/f_\pi) \Big]^n
	-\frac{3}{2}\frac{g_{3}^2}{(4\pi)^2} \frac{\mc{F}(m_\pi,\D_{\Xi^\prime\Xi},\mu)}{f_\pi^3}\, ,
\\ \nonumber\\
\label{eq:MXipC}
\frac{M_{\Xi^\prime_c}}{f_\pi} &= \frac{M_0+ \D_{\Xi^\prime\Xi}^{(0)}}{f_0}\frac{1}{1 + \d f(m_\pi/f_\pi)}
	- \frac{c_{\Xi^\prime}^r(\mu)}{4\pi}\frac{m_\pi^2}{f_\pi^2}%\Big[1 + \d f(m_\pi/f_\pi) \Big]^n
\nonumber\\&\qquad	
	-\frac{1}{2}\frac{g_{3}^2}{(4\pi)^2} \frac{\mc{F}(m_\pi,-\D_{\Xi^\prime\Xi},\mu)}{f_\pi^3}
	+\frac{1}{2}\frac{g_{2}^2}{(4\pi)^2} \frac{\mc{F}(m_\pi,0,\mu)}{f_\pi^3}\, .
\end{align}
%%%%%%%%%%%%%%%%
%%%  Lam Sig Fits
%%%%%%%%%%%%%%%%
\begin{table}[tb]
\begin{ruledtabular}
\begin{tabular}{c|c|ccccc|ccc}
Fit Range& $\D_{\Xi^\prime\Xi}/f_\pi^{\rm phys}$& $M_0 / f_\pi^{\rm phys}$& $c_\Xi^r(f_\pi)$& $c_{\Xi^\prime}^r(f_\pi)$& $g_{2}^2$& $g_{3}^2$& $\chi^2$& dof& $Q$\\
\hline
\texttt{m007}--\texttt{m030}&
	0.85(6)& 19.4(2)& 0.6(6)& 1.3(1.2)& 5.9(3.9)& $-$1.0(6)& 0.04& 3& 1.00\\
%\texttt{m007}--\texttt{m030}& 1&
%	0.85(6)& 19.4(2)& 0.5(5)& 1.4(1.2)& 7.6(5.3)& -1.1(8)& 0.01& 3& 1.00\\
\end{tabular}
\end{ruledtabular}
\caption{\label{tab:XiXipChPT}{Fit to $\Xi_c$ and $\Xi^\prime_c$ masses with NLO continuum formulae.}}
\end{table}
%%%%%%%%%%%%%%%%
%%%%%%%%%%%%%%%%

\begin{figure}[tb]
\begin{tabular}{c}
\includegraphics[width=0.5\textwidth]{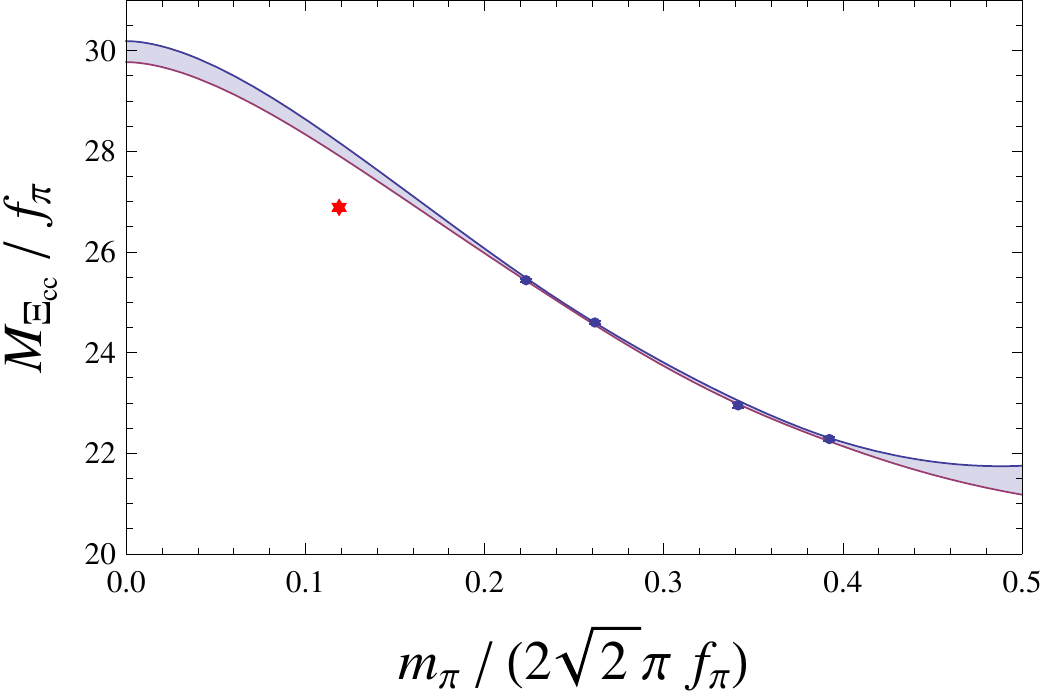}
\end{tabular}
\caption{\label{fig:XiCCChPT}NLO HH$\chi$PT extrapolation of $M_{\Xi_{cc}}$.}
\end{figure}

The masses of the remaining $J=1/2$ charmed baryons, $M_{\Xi_{cc}}$, $M_{\O_c}$ and $M_{\O_{cc}}$, can be treated independently.  The extrapolation formula for $M_{\Xi_{cc}}$ is similar to that of $M_{\S_c}$.  There is an axial coupling $g_{\Xi_{cc}\Xi_{cc}\pi}$ as well as $g_{\Xi^*_{cc}\Xi_{cc}\pi}$ where the second coupling is the axial transition coupling of the $J=3/2$ to the $J=1/2$-$\pi$ state.  The heavy quark symmetry also requires these couplings to be the same in the heavy quark limit.  At this order, we can treat the $J=3/2$ $\Xi_{cc}^*$ as degenerate with the $\Xi_{cc}$. The results are collected in Table~\ref{tab:XiCCChPT} and displayed in Figure~\ref{fig:XiCCChPT}, with the extrapolation formula~\cite{Mehen:2006vv}
\begin{align}
\frac{M_{\Xi_{cc}}}{f_\pi} =&\ \frac{M_0}{f_0}\frac{1}{1+\d f(m_\pi/f_\pi)}
	-\frac{c^{r}_{\Xi_{cc}}(\mu)}{4\pi}\frac{m_\pi^2}{f_\pi^2}%\Big[1 + \d f(m_\pi/f_\pi) \Big]^n
%\nonumber\\&\
	-\frac{g^2}{(4\pi)^2} \frac{\mc{F}(m_\pi,0,\mu)}{f_\pi^3}\, ,
%		\left[ \frac{\mc{F}(m_\pi,0,\mu)}{f_\pi^3} + \frac{\mc{F}(m_\pi,\D_{\Xi^*\Xi},\mu)}{f_\pi^3} \right]\, ,
\end{align}
where we have set $\D_{\Xi^*\Xi} = 0$ in this analysis, valid at this order in the heavy-quark expansion.
%%%%%%%%%%%%%%%%
%%%  Lam Sig Fits
%%%%%%%%%%%%%%%%
\begin{table}[tb]
\begin{ruledtabular}
\begin{tabular}{c|ccc|ccc}
Fit Range& $M_0 / f_\pi^{\rm phys}$& $c_{\Xi_{cc}}^r(f_\pi)$& $g^2$& $\chi^2$& dof& $Q$\\
\hline
\texttt{m007}--\texttt{m030}& 28.1(2)& 1.4(1.0)& $-$1.7(1.0)& 3.0& 1& 0.08\\
%\texttt{m007}--\texttt{m030}& 28.1(2)& 1.5(1.0)& -2.4(1.4)& 2.94& 1& 0.09\\
\end{tabular}
\end{ruledtabular}
\caption{\label{tab:XiCCChPT}Fit to $J=1/2$ $\Xi_{cc}$ mass with the NLO continuum heavy-hadron formula.}
\end{table}
%%%%%%%%%%%%%%%%
%%%%%%%%%%%%%%%%
One feature which is more pronounced in this fit is $g^2 < 0$.  Taken at face value, this would suggest the Lagrangian was non-Hermitian, and the theory not sensible.  Therefore, even though these fits reproduce the lattice data well and predict a mass within a few percent of the physical value, they must be taken with caution.  Most likely, as with the nucleon mass~\cite{WalkerLoud:2008pj}, there is a delicate cancelation of terms at different orders, and therefore one does not have confidence in these determinations of the LECs.

\begin{figure}[tb]
\begin{tabular}{cc}
\includegraphics[width=0.5\textwidth]{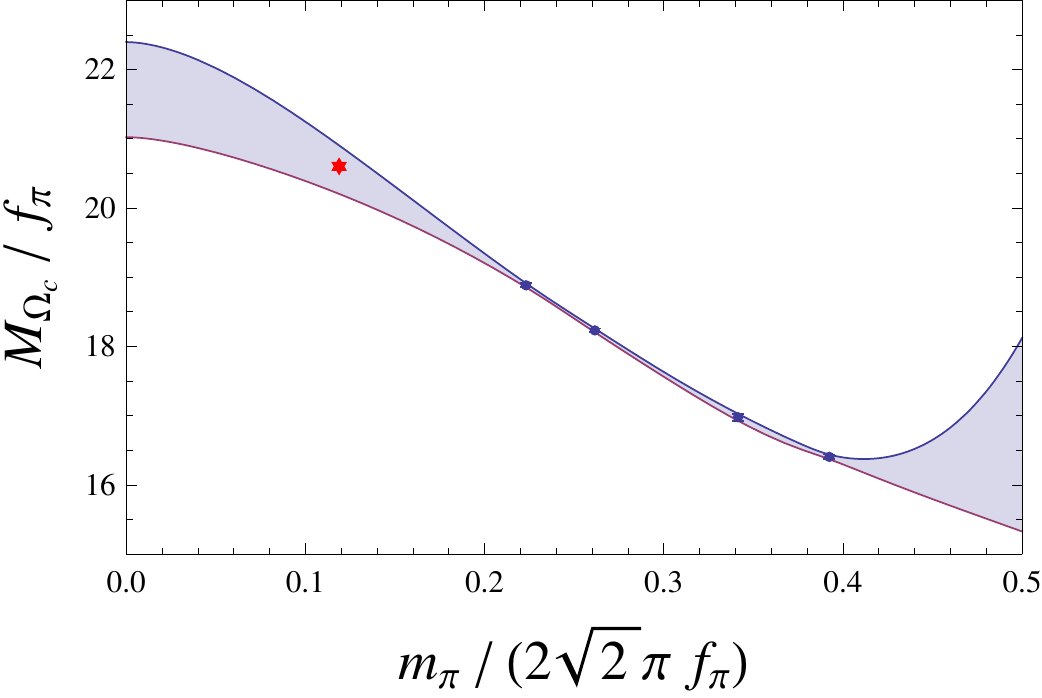}
&
\includegraphics[width=0.5\textwidth]{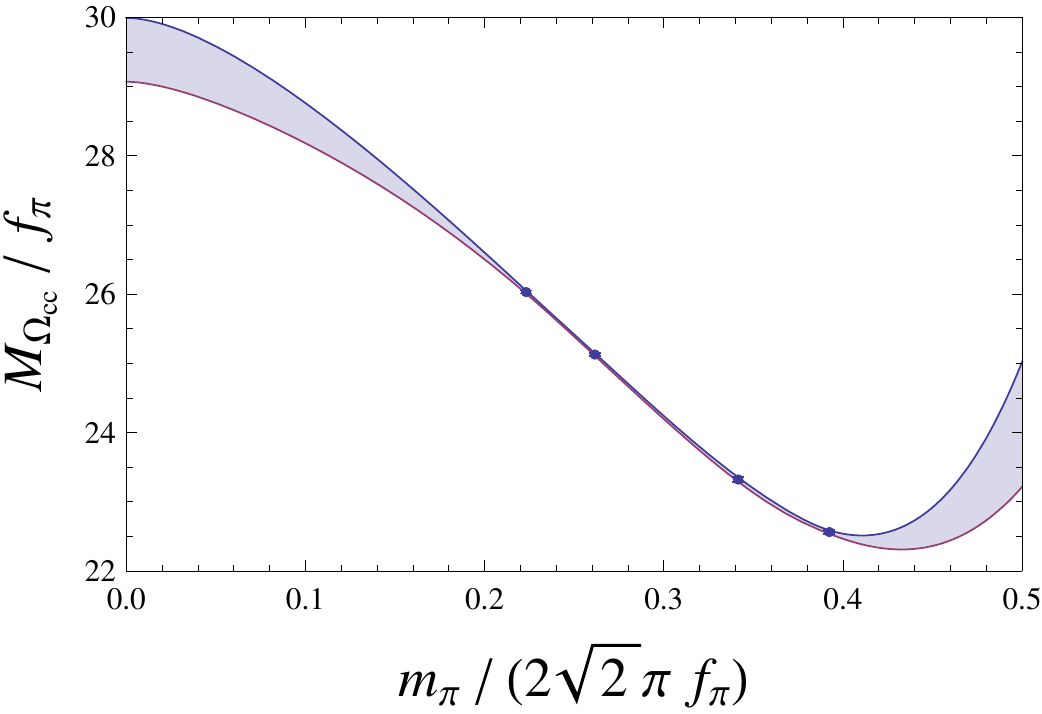}
\\
(a) & (b)
\end{tabular}
\caption{\label{fig:OmegaCCChPT}NLO HH$\chi$PT extrapolation of $M_{\O_c}$ (a) and $M_{\O_{cc}}$ (b).}
\end{figure}

Similar to the $s=-3$ $\O$, the $J=1/2$ $\O_c$ and $\O_{cc}$ do not have mass corrections which scale as $m_\pi^3$.  This is because these baryons do not contain any valence up or down quarks, and therefore, the leading $SU(2)$ axial coupling vanishes~\cite{Tiburzi:2004rh,Tiburzi:2008bk}.  The $SU(2)$ chiral extrapolation formula for these baryon masses is then expected to be as convergent as that for pions.  The mass extrapolation formula for the $\O_c$ and $\O_{cc}$ are both given by
\begin{align}
\frac{M_{\O}}{f_\pi} =&\ \frac{M_0}{f_0}\frac{1}{1+\d f(m_\pi/f_\pi)}
	-\frac{c^{r}_{\Xi_{cc}}}{4\pi}\frac{m_\pi^2}{f_\pi^2}%\Big[1 + \d f(m_\pi/f_\pi) \Big]^n
%\nonumber\\&
	+\frac{m_\pi^4}{(4\pi)^3 f_\pi^4} \left[ \alpha_\O^{(4)} \ln \left( \frac{m_\pi^2}{\mu^2} \right) + \beta_\O^{(4)(\mu)} \right]\, .
\end{align}
At this order, the two-loop corrections from $f_\pi$ should be included as corrections to $\a_\O^{(4)}$ and $\b_\O^{(4)}$.  Further, there is a $\ln^2(m_\pi)$ correction with fixed coefficient. However, since we only have four mass points, we cannot judge the quality of the fit anyway, so we ignore these corrections.  The results are collected in Table~\ref{tab:OmegaCCChPT} and displayed in Figure~\ref{fig:OmegaCCChPT}.  Performing a fit with $\a_\O = 0$ and $\b_\O = 0$ returns consistent mass predictions with smaller uncertainties.  We take the zero-degree-of-freedom fit as our central result as it provides a more conservative uncertainty.
%%%%%%%%%%%%%%%%
%%%  Omega C CC
%%%%%%%%%%%%%%%%
\begin{table}[tb]
\begin{ruledtabular}
\begin{tabular}{cc|cccc|ccc}
$\O$& Fit Range& $M_0 / f_\pi^{\rm phys}$& $c_{\O_{c}}^r(f_\pi)$& $\a_\O^{(4)}$& $\b_\O^{(4)}$& $\chi^2$& dof& $Q$\\
\hline
$\O_c$& \texttt{m007}--\texttt{m030}& 20.4(6)& $-$3.0(4.6)& 46(61)& $-$164(227)& 0.00& 0& -- \\
$\O_{cc}$& \texttt{m007}--\texttt{m030}& 27.7(4)& $-$7.3(3.0)& 109(40)& $-$392(149)& 0.00& 0& -- \\
\end{tabular}
\end{ruledtabular}
\caption{\label{tab:OmegaCCChPT}{Fit to $J=1/2$ $\O_{c}$ and $\O_{cc}$ masses with NLO continuum heavy-hadron formulae.}}
\end{table}
%%%%%%%%%%%%%%%%
%%%%%%%%%%%%%%%%

%%%%%%%%%%%%
%
%		POLYNOMIAL MPI EXTRAPOLATION
%
%%%%%%%%%%%%
\subsubsection{Polynomial Extrapolation}
\begin{figure}[tb]
\begin{tabular}{c}
\includegraphics[width=0.71\textwidth]{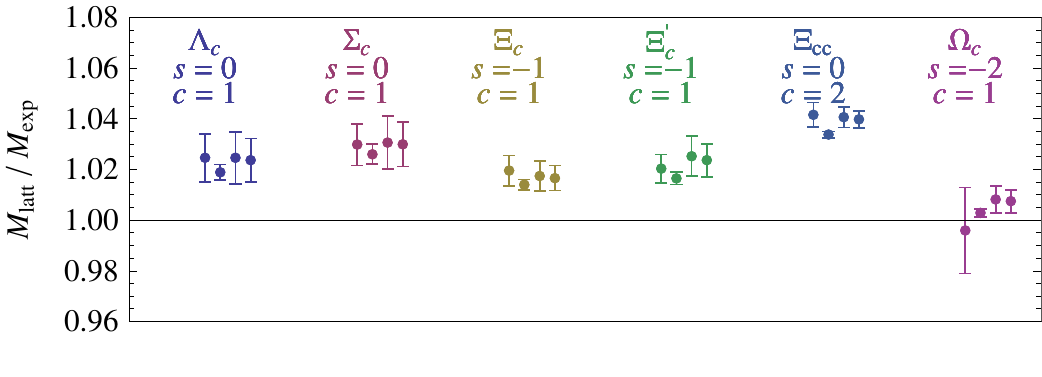}
\end{tabular}
\caption{\label{fig:RatioMB}Ratio of extrapolated masses to experimentally measured masses.  The first point represents the HH$\chi$PT fit, the second point is a fit with Eq.~\eqref{eq:Poly2}, the third with Eq.~\eqref{eq:Poly3} and the fourth with Eq.~\eqref{eq:Poly4}.}
\end{figure}

Given the issues of performing the heavy-hadron chiral extrapolations as discussed above, we also perform polynomial extrapolations in $m_\pi^2$.  We use the difference between the polynomial extrapolations and the heavy-hadron chiral extrapolations as an additional estimate of systematic extrapolation uncertainty.  We use up to three different polynomial fit functions for each of the charmed hadron masses:
\begin{align}
\frac{M_2}{f_\pi} &= \frac{M_0}{f_0}\frac{1}{1+\d f(m_\pi/f_\pi)}
	+c_2\frac{m_\pi^2}{f_\pi^2}\, , \label{eq:Poly2}\\
\frac{M_3}{f_\pi} &= \frac{M_0}{f_0}\frac{1}{1+\d f(m_\pi/f_\pi)}
	+c_2\frac{m_\pi^2}{f_\pi^2}
	+c_3 \frac{m_\pi^3}{f_\pi^3}\, ,\label{eq:Poly3}\\
\frac{M_4}{f_\pi} &= \frac{M_0}{f_0}\frac{1}{1+\d f(m_\pi/f_\pi)}
	+c_2\frac{m_\pi^2}{f_\pi^2}
	+c_4 \frac{m_\pi^4}{f_\pi^4}\, \label{eq:Poly4}.
\end{align}
In Figure~\ref{fig:RatioMB}, we display the results of these fits as well the heavy-hadron $\chi$PT fits as ratios with respect to the experimental masses. The experimental values for the baryon masses are taken from the Particle Data Group~\cite{Amsler:2008zzb}. As it can be seen, there is very little variation in the results of the extrapolated masses.  In all cases, the different extrapolations are consistent within one sigma.
%%%%%%%%%%%%%%%%
%%%%%%%%%%%%%%%%

In Table~\ref{tab:DirectMasses}, we provide the extrapolated baryon masses, taking the central value from the HH$\chi$PT extrapolations.  The first uncertainty is statistical and the second uncertainty is a comprehensive systematic uncertainty.  This systematic uncertainty is derived by comparing the polynomial light quark mass extrapolations to the HH$\chi$PT extrapolation.  Further, it includes the uncertainty associated with the choice of fitting window for the correlators as well. Except for the $\O_{c}$, the extrapolated masses are systematically high, indicative of a discretization error.

%%%%%%%%%%%%%%%%
%%%  DIRECT MASS RESULTS
%%%%%%%%%%%%%%%%
\begin{table}[tb]
\begin{ruledtabular}
\begin{tabular}{c|ccccccc}
State & $M_{\L_c}$& $M_{\Xi_c}$& $M_{\S_c}$& $M_{\Xi^\prime_c}$& $M_{\O_c}$& $M_{\Xi_{cc}}$& $M_{\O_{cc}}$ \\
($J=1/2$)& [{\rm MeV}]& [{\rm MeV}]& [{\rm MeV}]& [{\rm MeV}]& [{\rm MeV}]& [{\rm MeV}] & [{\rm MeV}] \\
\hline
Prediction& 2342(22)(11)& 2527(17)(13)& 2527(20)(08)& 2638(17)(10)& 2687(46)(16)& 3665(17)(14)& 3680(31)(38) \\
Exp. Mass & 2286& 2468& 2454& 2576& 2698& 3519& --\\
\end{tabular}
\end{ruledtabular}
\caption{\label{tab:DirectMasses}Direct light/heavy quark mass extrapolation of the $J=1/2$ charmed baryon spectrum.}
\end{table}
%%%%%%%%%%%%%%%%%%%%%%%%%%%

%%%%%%%%%%%%
%
%		MASS SPLITTINGS AND DISCRETIZATION ERRORS
%
%%%%%%%%%%%%
\subsection{Discretization Errors and Mass Splittings\label{sec:MSplitt}}
In this work, we have performed calculations at only a single value of the lattice spacing, with $a \sim 0.125$~{fm}, prohibiting us from performing a continuum extrapolation.  However, we can take advantage of various symmetries and power counting to make a reasonable estimate of the discretization errors present in our calculation.%
%FOOTNOTE
\footnote{With a single lattice spacing, we can not disentangle both the discretization errors and the tuning of the charm quark mass.  The effects we discuss here as discretization errors are really a combination of the two.} 
In these heavy-light systems, the discretization errors arise both from the light and heavy quark actions.  The corrections from both generically scale as $\mc{O}(a^2)$ for each of the charmed baryon masses.  If we consider $SU(3)$ symmetry, then the leading discretization errors for all baryons in a given $SU(3)$ multiplet must be the same, with corrections scaling as $\mc{O}(a^2 (m_s - m_u))$.  Further, if one considers the combined large-$N_c$, $SU(3)$ and heavy-quark symmetries~\cite{Jenkins:1996de}, then all the singly charmed baryon masses we calculate in this work share a common discretization correction to their masses, with sub-leading corrections scaling as $\mc{O}(a^2 / N_c)$ as well as the $SU(3)$ breaking corrections.  Therefore, all the singly charmed baryon masses we compute in this work, $\{\L_c, \Xi_c, \S_c, \Xi^\prime_c, \O_c\}$ share a common discretization correction, which happens to be the dominant discretization error.  The same analysis holds for the doubly charmed baryons as well, $\{\Xi_{cc},\O_{cc}\}$ with a common error, albeit different from the singly charmed correction.%
%FOOTNOTE
\footnote{With the full $J=3/2$ and $J=1/2$ heavy baryon mass spectrum, one could perform an analysis of the large-$N_c$ baryon mass relations~\cite{Dashen:1993jt,Jenkins:1995td} as has recently been performed for the light quark octet and decuplet baryons~\cite{Jenkins:2009wv}.} 
It is therefore advantageous to consider extrapolations of baryon mass splittings, as these mass splittings exactly cancel the leading discretization errors.

Before proceeding with the analysis of the mass splittings, we first use power counting arguments to estimate the discretization errors.  The leading discretization corrections from the light and heavy quark actions can be estimated as~\cite{Kronfeld:2003sd}
\begin{align}
	&\d_q(a^2) = \frac{1}{2}(a p)^2 \L_{QCD}\, ,&
\nonumber\\
	&\d_Q(a^2) = \frac{\a_s(m_c) (ap)}{2(1 + a m_c)} \L_{QCD}\, ,&
\end{align}
where $p$ is a typical momentum scale, of the order of $\L_{QCD}$, the characteristic hadronic scale.  To be conservative, we can take $\L_{QCD} = 700$~MeV which leads to the estimates
\begin{align}
	&\d_q(a^2) = 68 \textrm{ MeV}\, ,&
\nonumber\\
	&\d_Q(a^2) = 19 \textrm{ MeV}\, .&
\end{align}
When considering mass splittings amongst a given $SU(3)$ multiplet, these leading errors become further suppressed by $m_s - m_u$ effects, 
\begin{align}
	&\d \D M_q(a^2) = \frac{1}{2}(a p)^2 \L_{QCD} \frac{m_K^2 - m_\pi^2}{\L_{\chi}^2} \, ,&
\nonumber\\
	& \d \D M_Q(a^2) = \frac{\a_s(m_c) (ap)}{2(1 + a m_c)} \L_{QCD} \frac{m_K^2 - m_\pi^2}{\L_{\chi}^2} \, .&
\end{align}
Mass splittings between the two singly charmed $SU(3)$ multiplets, $\D M^{6,\bar{3}}$, would receive similar discretization corrections, with the extra suppression of $1/N_c$.
Combining these estimates in quadrature,%
%FOOTNOTE
\footnote{For the doubly charmed baryon masses, we double the estimated heavy quark discretization error.  As mentioned above, this uncertainty also includes any miss-tuning of the charm quark mass, and thus a double charmed baryon will be miss-tuned twice as much.} 
 we estimate the discretization errors for the baryon masses, and various mass splittings (using $\L_{\chi} = 2\sqrt{2}\pi f_\pi$ and the physical kaon and pion masses)
\begin{align}\label{eq:discoErrors}
	&\d M_{h_c} = 71 \textrm{ MeV}\, ,&
\nonumber\\
	&\d M_{h_{cc}} = 78 \textrm{ MeV}\, ,&
\nonumber\\
	&\d\D M_{h_c} = 12 \textrm{ MeV}\, ,&
\nonumber\\
	&\d\D M_{h_{cc}} = 13 \textrm{ MeV}\, ,&
\nonumber\\
	&\d\D M^{6,\bar{3}}_{h_c} = 24 \textrm{ MeV}\, ,&
\nonumber\\
	&\d\D M^{6,\bar{3}}_{h_{cc}} = 26 \textrm{ MeV}\, .&
\end{align}

%%%%%%%%%%%%%%%%
\begin{figure}[tb]
\begin{tabular}{cc}
\includegraphics[width=0.5\textwidth]{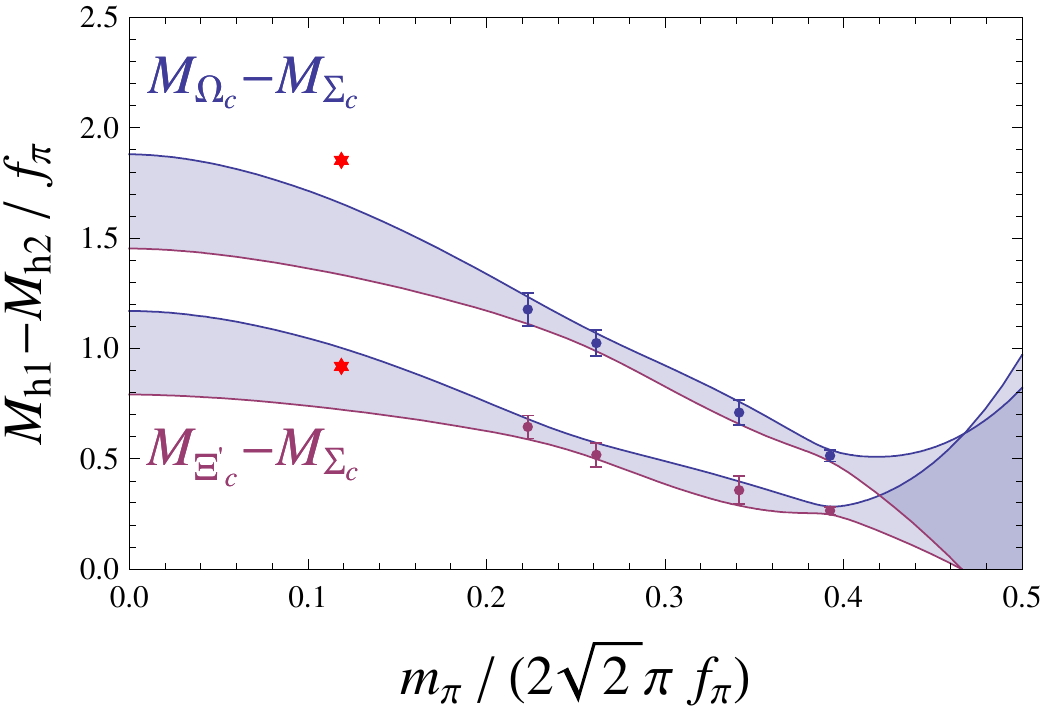}&
\includegraphics[width=0.5\textwidth]{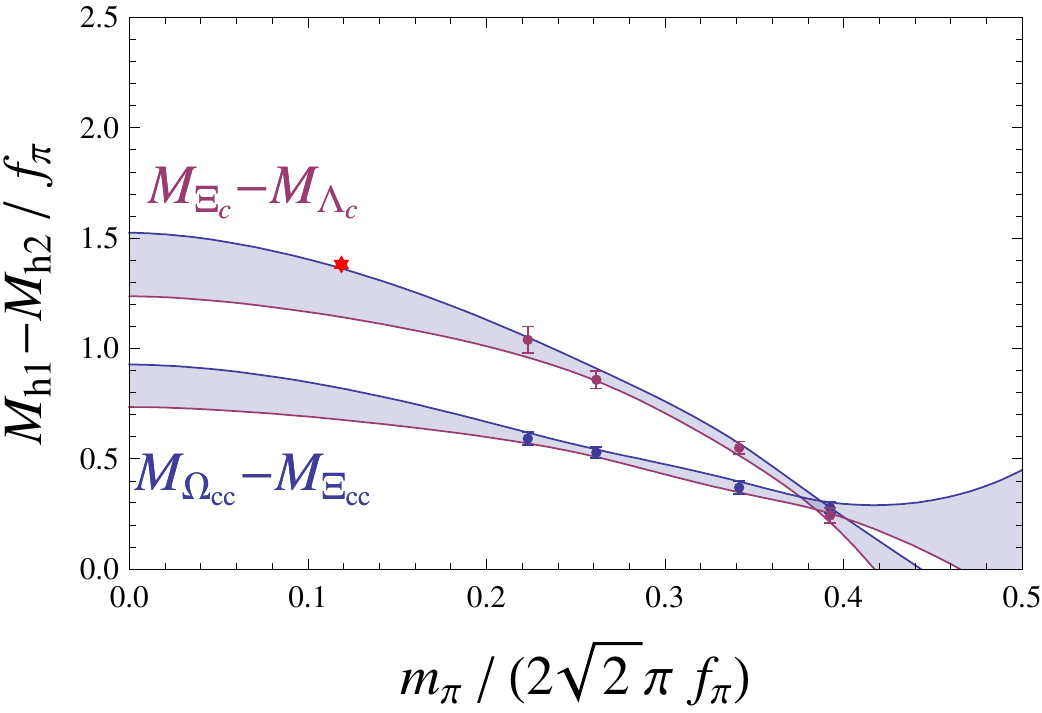}
\end{tabular}
\caption{\label{fig:MassSplitExtrap}Polynomial extrapolations of of $J=1/2$ mass splittings amongst heavy-quark--$SU(3)$ multiplets with Eq.~\eqref{eq:Poly4}.}
\end{figure}
\begin{figure}[tb]
\begin{tabular}{c}
\includegraphics[width=0.7\textwidth]{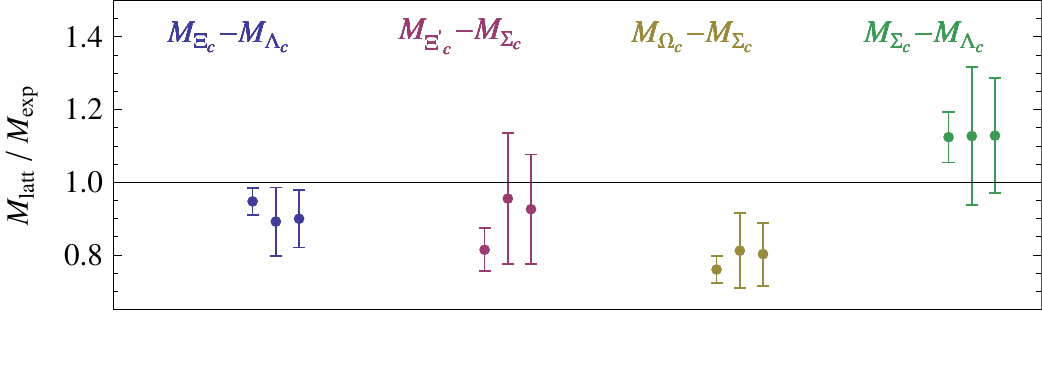}
\end{tabular}
\caption{\label{fig:RatioMBSplitt}Ratio of extrapolated mass splittings to experiment~\cite{Amsler:2008zzb}.  The first point is a fit with Eq.~\eqref{eq:Poly2}, the second with Eq.~\eqref{eq:Poly3} and the third with Eq.~\eqref{eq:Poly4}.}
\end{figure}

%%%%%%%%%%%%%%%%
Given our limited number of light-quark mass values, we are not able to perform the (mixed-action) HH$\chi$PT analysis of the mass splittings.
We therefore perform our fits using the polynomial fit functions, Eqs.~\eqref{eq:Poly2}--\eqref{eq:Poly4}, with $M_0$ replaced by $\D_{h_2 h_1}^{(0)}$.  We perform the extrapolations of the mass splittings, $M_{\Xi_c} - M_{\L_c}$, $\{M_{\Xi^\prime_c},M_{\O_c}\} - M_{\S_c}$, $M_{\S_c} - M_{\L_c}$ and $M_{\O_{cc}} - M_{\Xi_{cc}}$.
In Figure~\ref{fig:MassSplitExtrap} we display the extrapolation of these mass splittings using Eq.~\eqref{eq:Poly4} and in Figure~\ref{fig:RatioMBSplitt} we show the ratio of these fits to the experimental values.  Our final predicted splittings are determined by using the quartic fit function as the central value with the differences from the quadratic and cubic fits to estimate light quark mass extrapolation errors (in addition to those from the quartic fit).
%%%%%%%%%%%%%%%%
%%%%%%%%%%%%%%%%

As discussed earlier in this section, the dominant discretization error in the mass calculations is common to all baryons, given the various symmetries.  Therefore, this correction will shift all the baryon masses in one direction.  
We can determine the sign of this correction in the following manner.  First, we can determine the singly charmed baryon spectrum by taking our extrapolated mass splittings, column $(a)$ of Table~\ref{tab:results}, and using $M_{\L_c}^{\rm phys}$ and $M_{\S_c}^{\rm phys}$ as reference scales, $M_{h_c}^{\rm split} = M_{\L_c,\S_c}^{\rm phys} + \D M_{h_c-\L_c,\S_c}$, resulting in the predicted masses, Table~\ref{tab:results} $(b)$.  We then compare these to our direct mass extrapolations $M_{h_c}^{\rm direct}$, given in Table~\ref{tab:DirectMasses}.  The first method is free of the leading discretization errors while the second is not.  We can then construct the quantity,
\begin{align}
\d M_c(a^2) = \frac{1}{N_{h_c}} 
	\sum_{h_c} \left( M_{h_c}^{\rm direct} - M_{h_c}^{\rm split}\right)\, ,
\end{align}
which is a measure of these discretization errors.  The sum runs over all four singly charmed baryons $h_c$ for which we have both methods to determine the masses ($N_{h_c}=4$).  The first thing to note is that every element contributing to the sum is a positive quantity, suggesting the discretization errors increase the baryon masses.  It is also interesting to note that in our calculation, $\d M_c(a^2) = 59$~{MeV}, comparable to our estimated leading discretization effects, Eq.~\eqref{eq:discoErrors}.
We can then refine our estimate of the leading discretization errors to be
\begin{align}
	&\d M_{h_c} = {}^{+0}_{-71} \textrm{ MeV}\, ,&
\nonumber\\
	&\d M_{h_{cc}} = {}^{+0}_{-78} \textrm{ MeV}\, ,&
\end{align}
where we have also assumed that the doubly charmed discretization errors do not change sign relative to the singly charmed baryon corrections.  Our final numbers, collected in Table~\ref{tab:results}, include these discretization error estimates in the quoted uncertainties.

%%%%%%%%%%%%%%%%
%%%  MASS RESULTS
%%%%%%%%%%%%%%%%
\begin{table}[tb]
\begin{ruledtabular}
\begin{tabular}{c|c|c||c|c|c|c}
State& Latt. Pred.& Exp.& State & Mass Split.& Direct Mass&Exp. Mass \\
& [{\rm MeV}] & [{\rm MeV}] &&  [{\rm MeV}] &  [{\rm MeV}] &  [{\rm MeV}] \\
\hline
&&
	&$M_{\L_c}$& 
	&$2342\pm22\pm11\,{}^{+0}_{-71}$& 2286 \\
$M_{\Xi_c} - M_{\L_c}$& $164\pm14\pm23\pm12$& 182
	&$M_{\Xi_c}$& $2450\pm14\pm23\pm12$
	&$2527\pm17\pm13{}^{+0}_{-71}$ &2468\\
\cline{4-7}
$M_{\S_c} - M_{\L_c}$& $190\pm27\pm18\pm27$&168
	&$M_{\S_c}$
	&$2476\pm27\pm18\pm27$
	&$2527\pm20\pm8\, {}^{+0}_{-71}$ &2454 \\
$M_{\Xi^\prime_c} - M_{\S_c}$& $113\pm18\pm8\pm12$& 122
	&$M_{\Xi^\prime_c}$& $2567\pm18\pm8\pm12$
	&$2638\pm17\pm10\, {}^{+0}_{-71}$ &2576 \\
$M_{\O_c} - M_{\S_c}$& $195\pm21\pm7\pm12$& 244
	&$M_{\O_c}$& $2649\pm21\pm7\pm12$
	&$2687\pm46\pm16\, {}^{+0}_{-71}$ &2698 \\
\cline{4-7}
&&
	&$M_{\Xi_{cc}}$
	& 
	& $3665\pm17\pm14\,{}^{+0}_{-78}$ &3519\\
$M_{\O_{cc}} - M_{\Xi_{cc}}$& $98\pm9\pm22\pm13$& --
	&$M_{\O_{cc}}$& $3763\pm19\pm26\, {}^{+13}_{-79}$
	&$3680\pm31\pm38\, {}^{+0}_{-78}$ &--\\
\end{tabular}
\end{ruledtabular}
\begin{tabular}{ccc}
\qquad\qquad\qquad\qquad$(a)$\qquad\qquad \qquad \qquad \qquad \qquad
& \qquad\qquad$(b)$\qquad \qquad
	& \qquad\qquad \qquad$(c)$\qquad \qquad \qquad \qquad
\end{tabular}
\caption{\label{tab:results}Resulting charmed spectrum, extrapolated in the light-quark mass to the physical $m_\pi^{\rm phys} / f_\pi^{\rm phys}$ point.  In $(a)$ we display the mass splittings of the baryons related by $SU(3)$ and large $N_c$ symmetry.  As discussed in detail in the text, the first uncertainty is statistical, the second is systematic and the third is our estimate of discretization errors.  These are the central results of this work.
In $(b)$, we display our resulting baryon spectrum determined using the experimental values of $M_{\L_c}^{\rm exp}$ and $M_{\S_c}^{\rm exp}$, combined with our splittings in $(a)$.  For the $\O_{cc}$, we use our extrapolated value of $M_{\Xi_{cc}}$ given the present uncertainty in the experimental value.
In $(c)$, we present the results of our direct mass extrapolations, including our estimated discretization errors.  The results from the two methods are consistent at the one-sigma level.}
\end{table}
%%%%%%%%%%%%%%%%%%%%%%%%%%%

%%%%%%%%%%%%
%
%		DISCUSSION AND CONCLUSION
%
%%%%%%%%%%%%
\section{Discussion and Conclusions\label{sec:results}}

\begin{figure}
\includegraphics[width=0.65\textwidth]{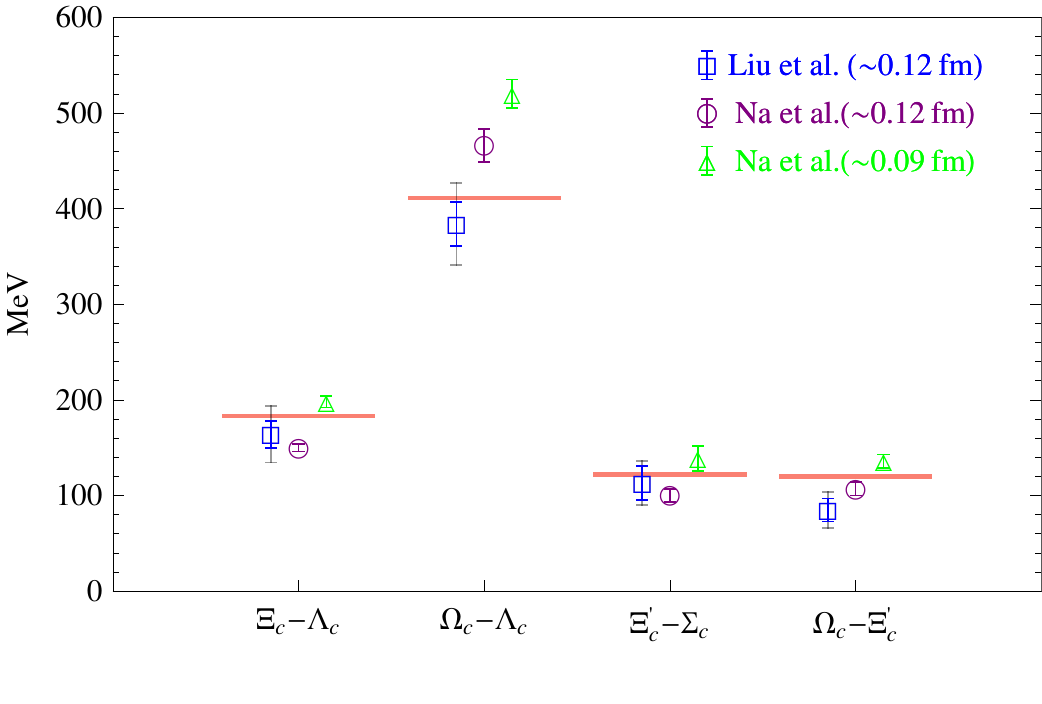}
\caption{\label{fig:splittings} Comparison among charmed baryon mass splittings of dynamical lattice calculations.  The results of Na et al. are taken from Ref.~\cite{Na:2008hz}.}
\end{figure}

The central results of this work are the predicted mass splittings, displayed in the left panel of Table~\ref{tab:results}.  The first uncertainty is statistical and the second uncertainty is a comprehensive systematic as discussed in the text.  The third uncertainty is an estimate of discretization errors, which must scale as $\mc{O}(a^2(m_s - m_u))$ for members of the same $SU(3)$ multiplet or $\mc{O}(a^2 / N_c)+\mc{O}(a^2(m_s - m_u))$ otherwise, as dictated by the approximate symmetries.
These results have been extrapolated to the physical charm quark mass and the physical light quark mass defined respectively by
\begin{align*}
\frac{M_{\eta_c}^{\rm phys} + 3 M_{J/\Psi}^{\rm phys}}{4 f_\pi^{\rm phys}} &= 23.47\,,\\
\frac{m_\pi^{\rm phys}}{f_\pi^{\rm phys}} &= 1.056\, .
\end{align*}
To perform these extrapolations, we first formed the dimensionless ratios $(M_{h_1}^{\rm latt} - M_{h_2}^{\rm latt})/f_\pi^{\rm latt}$, taking into account the known light-quark mass dependence of $f_\pi$.  The mass splittings in {MeV} are then determined with $f_\pi = 130.7$~{MeV}.  These physical values are all taken from the PDG~\cite{Amsler:2008zzb}.
In Fig.~\ref{fig:splittings}, we compare some of our mass splitting results with those of Gottlieb and Na~\cite{Na:2007pv,Na:2008hz}, the only other dynamical calculation of the charmed baryon spectrum.  They used the same MILC gauge ensembles, as well as the fine $a\sim 0.09$~fm lattices.  For the light quark propagators, they used staggered fermions, and for the heavy quark, an interpretation of the Fermilab action was used, defining the charm mass with the kinetic mass instead of the rest mass.  Their work is still somewhat preliminary and does not yet provide a systematic uncertainty.  However, our results are consistent with theirs, especially those on the same ensembles with $a\sim 0.125$~{fm}.
%%%%%%%%%%%%%%%%%%%%%%%%%%%
%				FIGURE:  LATTICE MASS SPLITTING COMPARISON
%%%%%%%%%%%%%%%%%%%%%%%%%%%
%%%%%%%%%%%%%%%%%%%%%%%%%%%

\begin{figure}
\includegraphics[width=0.65\textwidth]{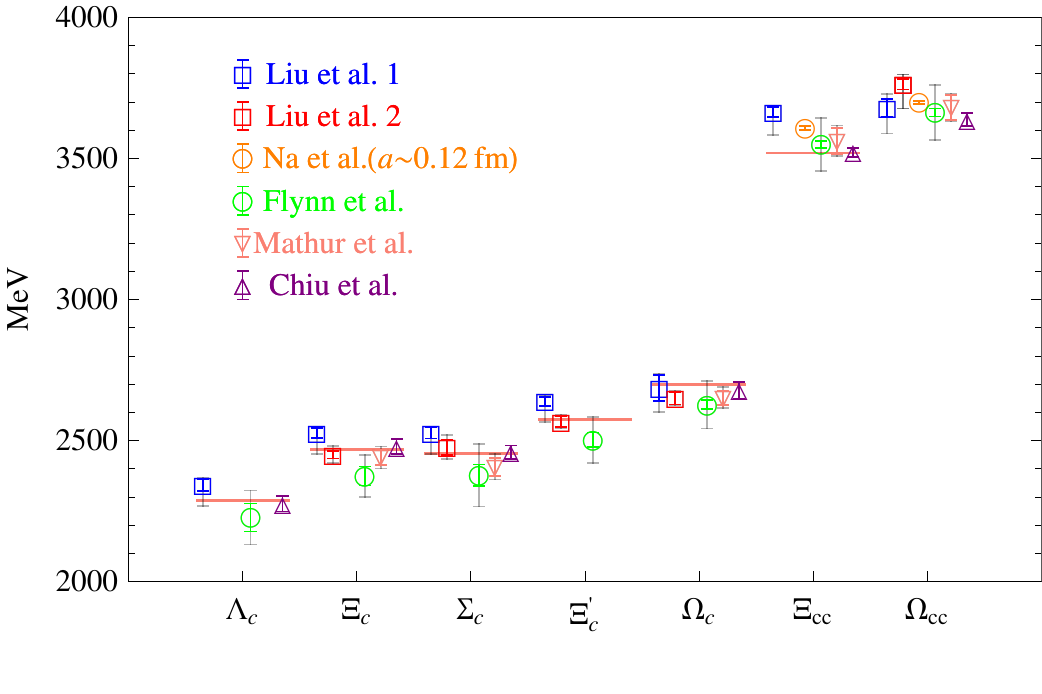}
\caption{\label{fig:summary_charmedB} A summary of charmed baryon masses in {MeV} calculated using LQCD.
We show both of our methods for obtaining the spectrum, the direct mass extrapolation (Liu et al. 1) and also using the extrapolated mass splittings, combined with $M_{\L_c}^{\rm exp}$ and $M_{\S_c}^{\rm exp}$ (Liu et al. 2).  These results are taken from Table~\ref{tab:results}. 
The other results, displayed for comparison, are taken from Table~\ref{tab:charmedB_summary}.}
\end{figure}

We additionally use these mass splittings, combined with the experimental value of $M_{\L_c}^{\rm exp}$ and $M_{\S_c}^{\rm exp}$ to determine the $J=1/2$ baryon masses.
 Aside from the $\Xi_{cc}$ state,%
 %FOOTNOTE
\footnote{Because the $\Xi_{cc}$ has not been verified by multiple experimental groups~\cite{Mattson:2002vu,Ocherashvili:2004hi,Aubert:2006qw,Chistov:2006zj,Amsler:2008zzb}, we chose to use our extrapolated value of $M_{\Xi_{cc}}$, combined with our extrapolated value of $M_{\O_{cc}} - M_{\Xi_{cc}}$ to make a prediction for the $\O_{cc}$ mass.} 
 the masses determined in this way are consistent with our direct mass extrapolation results, Table~\ref{tab:results}~$(c)$, after including our estimated discretization errors.  We used power counting arguments~\cite{Oktay:2008ex,Kronfeld:2003sd} to estimate the size of these corrections and we compared our two methods of determining the baryon masses to determine the expected sign of the leading discretization corrections.  In Fig.~\ref{fig:summary_charmedB}, we display our resulting mass calculations using the results from both the mass splitting method (Liu et al. 2) as well as the direct extrapolation of the masses (Liu et al. 1).  Additionally, we compare these with results from previous calculations, found in the Refs. of Table~\ref{tab:charmedB_summary} (for those calculations with more than one lattice spacing, we show only the results from the ensemble with lattice spacing closest to the one used in this work).
%%%%%%%%%%%%%%%%%%%%%%%%%%%
%		FIGURE: LATTICE MASS COMPARISON
%%%%%%%%%%%%%%%%%%%%%%%%%%%
%%%%%%%%%%%%%%%%%%%%%%%%%%%
%%%%%%%%%%%%%%%%%%%%%%%%%%%
\begin{table}
\begin{center}
\footnotesize
\begin{ruledtabular}
\begin{tabular}{c|ccccccc}
Group & $N_{\rm f}$ & $S_{\rm H}$ & $a_t^{-1}$ (GeV)   & $L$ (fm)\\
\hline
Bowler et~al.~\cite{Bowler:1996ws}    & 0   & tree clover~\cite{Sheikholeslami:1985ij} & 2.9   &  1.63 \\
Lewis et~al.~\cite{Lewis:2001iz}  & 0   & D234~\cite{Alford:1995dm} & 1.8, 2.2, 2.6   &  1.97\\
Mathur et~al.~\cite{Mathur:2002ce} & 0 & NRQCD~\cite{Sheikholeslami:1985ij} & 1.8, 2.2 & 2.64,2.1\\
Flynn et~al.~\cite{Flynn:2003vz} & 0 & NP clover  & 2.6 & 1.82\\
Chiu et~al.\cite{Chiu:2005zc} & 0 & ODWF~\cite{Chiu:2002ir} & 2.23 &1.77\\
Na et~al.\cite{Na:2007pv,Na:2008hz} & $2+1$ & Fermilab~\cite{ElKhadra:1996mp} & 2.2, 1.6, 1.3& 2.5\\
This work& $2+1$& Fermilab& 1.6& 2.5
\end{tabular}
\end{ruledtabular}
\caption{\label{tab:charmedB_summary}Summary of existing charmed baryon published calculations from lattice QCD. Please refer to the above references and references within for more details.}
\end{center}
\end{table}
%%%%%%%%%%%%%%%%%%%%%%%%%%%

\begin{figure}
\begin{tabular}{c}
\includegraphics[width=0.65\textwidth]{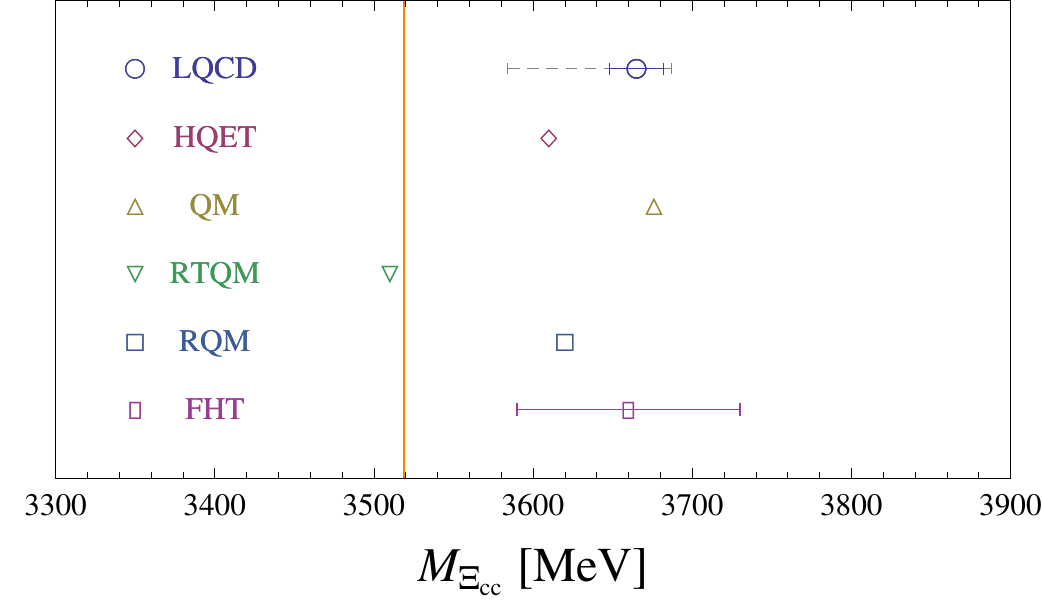}\\
\includegraphics[width=0.65\textwidth]{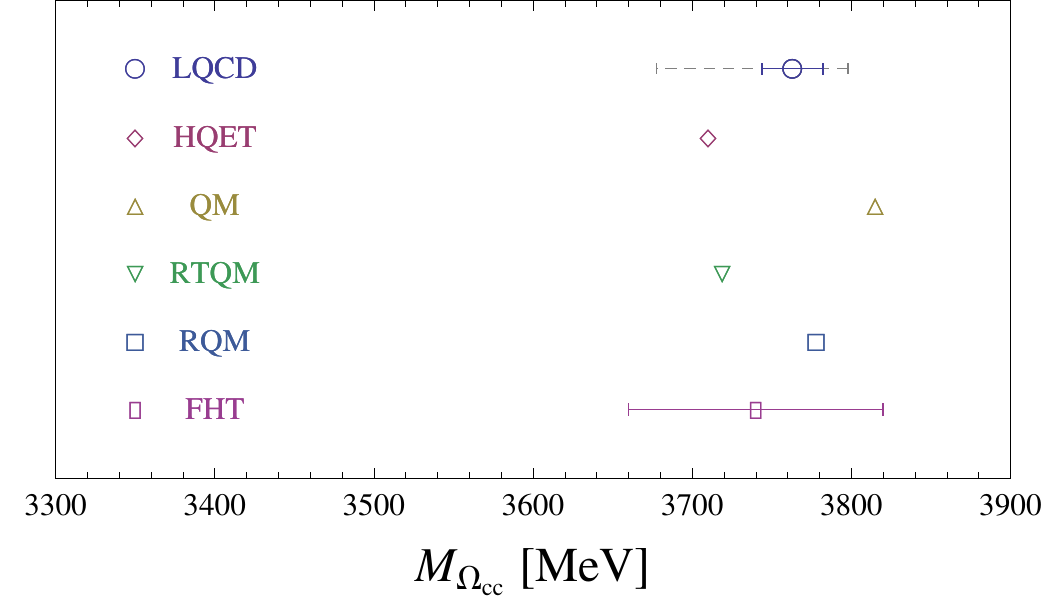}\\
\end{tabular}
\caption{\label{fig:DoublyCharmedB} Comparison of theoretical predictions for doubly charmed baryons of spin 1/2. ``LQCD'' is the lattice QCD calculation done in this work with solid error bars for the statistical error and dashed bars for the total error including the estimated systematic;
``QM'' is taken from a recent quark-model calculation~\cite{Roberts:2007ni};
``RTQM'' is the result of relativistic three-quark model~\cite{Martynenko:2007je};
``RQM'' and ``HQET'' are from the relativistic quark model~\cite{Ebert:2002ig} and the heavy-quark effective theory~\cite{Korner:1994nh} respectively;
note that there is no error estimation done in these calculations.
``FHT'' is based on the Feynman-Hellmann theorem~\cite{Roncaglia:1995az}, where rough uncertainties are estimated.}
\end{figure}

Finally, we compare the doubly charmed baryons with the predictions of theoretical models, as shown in Fig.~\ref{fig:DoublyCharmedB}. Although the SELEX Collaboration has reported the first observation of doubly charmed baryons, searches by the BaBar~\cite{Aubert:2006qw}, Belle~\cite{Chistov:2006zj} and Focus~\cite{Ratti:2003ez} Collaborations have not confirmed their results. This makes it interesting to look back to the theory to see where the various predictions lie.
We compare with a selection of other theoretical results, such as a recent quark-model calculation~\cite{Roberts:2007ni}, relativistic three-quark model~\cite{Martynenko:2007je}, the relativistic quark model~\cite{Ebert:2002ig}, the heavy quark effective theory~\cite{Korner:1994nh} and the Feynman-Hellmann theorem~\cite{Roncaglia:1995az}.
We compute the mass of $\Xi_{cc}$ to be $3665\pm17\pm14\,{}^{+0}_{-78}$~MeV, which is higher than what SELEX observed, although less than two sigma with our estimated discretization errors; most theoretical results suggest that the $\Xi_{cc}$ that is about 100--200~MeV higher than the SELEX experimental value.  To improve this situation, we need results at multiple lattice spacings to reduce this systematic uncertainty.
The $\Omega_{cc}$ mass prediction made by this work is $3763\pm19\pm26\,{}^{+13}_{-79}$~MeV, and the overall theoretical expectation is for the $\Omega_{cc}$ to be 3650--3850~MeV.  We hope that upcoming experiments will be able to resolve these mysteries of doubly charmed baryons.

Our largest uncertainty presently arises from the lack of a continuum extrapolation.  Therefore, in the future we plan to extend these calculations to a second lattice spacing.  This will hopefully allow us to significantly reduce the size of our discretization errors.  Additionally, we are extending our calculation to include the spin-3/2 spectroscopy.

\begin{acknowledgements}
AWL and KO would like to thank Brian Tiburzi for helpful discussions.  We would like to thank Heechang Na and Steve Gotlieb for providing us with their spectrum numbers from Ref.~\cite{Na:2008hz}. 
We thank the NPLQCD collaboration for sharing their propagators: most of the light-quark and all of the strange-quark propagators used in this work; we also thank LHPC for some light-quark propagators.  We would also like to thank Jo Dudek for a careful reading of our manuscript.
These calculations were performed using the Chroma software suite~\cite{Edwards:2004sx}, on computer clusters at Jefferson Laboratory (USQCD SciDAC supported) and the College of William and Mary (Cyclades cluster supported by the Jeffress Memorial Trust grant J-813).
LL and HL are supported by Jefferson Science Associates, LL under U.S. DOE Contract No. DE-AC05-06OR23177. HL is also supported by the U.S. Dept. of Energy under Grant No. DE-FG03-97ER4014.
KO is supported in part by the Jeffress Memorial Trust grant J-813, DOE OJI grant DE-FG02-07ER41527 and DOE grant DE-FG02-04ER41302.
AWL is supported under the U.S. DOE OJI grant DE-FG02-07ER-41527.
\end{acknowledgements}

\bibliographystyle{apsrev}
\bibliography{charmed}

\end{document}